\documentclass[12pt,english]{article}
\usepackage{amsmath}
\usepackage{amsthm}
\usepackage{mathptmx}
\usepackage{newtxmath}
\usepackage[T1]{fontenc}
\usepackage[latin9]{inputenc}
\usepackage{geometry}
\usepackage{color}
\usepackage{enumitem}
\usepackage{setspace}
\makeatletter
\IfFileExists{candleshape.def}{%
 \input{candleshape.def}}{}
\IfFileExists{dropshape.def}{%
 \input{dropshape.def}}{}
\IfFileExists{TeXshape.def}{%
 \input{TeXshape.def}}{}
\IfFileExists{triangleshapes.def}{%
 \input{triangleshapes.def}}{}

\newlength{\lyxlabelwidth}      
\@ifundefined{lettrine}{\usepackage{lettrine}}{}
\numberwithin{equation}{section}
\numberwithin{figure}{section}
\numberwithin{table}{section}
	\newenvironment{elabeling}[2][]%
	{\settowidth{\lyxlabelwidth}{#2}
		\begin{description}[font=\normalfont,style=sameline,
			leftmargin=\lyxlabelwidth,#1]}
	{\end{description}}
\theoremstyle{definition}
\ifx\thechapter\undefined
  \newtheorem{defn}{\protect\definitionname}
\else
  \newtheorem{defn}{\protect\definitionname}[chapter]
\fi
\theoremstyle{definition}
\ifx\thechapter\undefined
  \newtheorem{example}{\protect\examplename}
\else
  \newtheorem{example}{\protect\examplename}[chapter]
\fi
\theoremstyle{plain}
\ifx\thechapter\undefined
  \newtheorem{thm}{\protect\theoremname}
\else
  \newtheorem{thm}{\protect\theoremname}[chapter]
\fi
\theoremstyle{plain}
\ifx\thechapter\undefined
  \newtheorem{lem}{\protect\lemmaname}
\else
  \newtheorem{lem}{\protect\lemmaname}[chapter]
\fi

\makeatother

\usepackage{babel}
\providecommand{\definitionname}{Definition}
\providecommand{\examplename}{Example}
\providecommand{\lemmaname}{Lemma}
\providecommand{\theoremname}{Theorem}

\begin{document}
\title{How to De-reserve Reserves: Admissions to Technical Colleges in India\thanks{\emph{First version:} March, 2021. We are grateful to David Delacrétaz,
Yash Kanoria, Assaf Romm, and Rakesh Vohra for their detailed feedbacks.
We also thank Nick Arnosti, Joydeep Bhattacharya, Juan Carlos Cordoba,
Federico Echenique, Aram Grigoryan, Debasis Mishra, Tarun Sabarwal,
Arunava Sen, Rajesh Singh, Bumin Yenmez and the seminar audiences
at Indian Statistical Institute Delhi, Iowa State, and conference
participants of 2021 ACM EAAMO, 2021 Latin American Meeting of the
Econometric Society, 2021 European Winter Meeting of the Econometric
Society, 17th Matching in Practice, 2021 Winter School of the Econometric
Society at DSE, and Kansas CARE-Colloquium for helpful comments.}}
\author{Orhan Aygün\thanks{Aygün: Department of Economics, Bo\u{g}aziçi University (Email: orhan.aygun@boun.edu.tr).}
$\quad$and $\quad$Bertan Turhan\thanks{Turhan: Department of Economics, Iowa State University (Email: bertan@iastate.edu).}}
\date{May,  2022}
\maketitle
\begin{abstract}
We study the\emph{ joint} implementation of\emph{ reservation} and
\emph{de-reservation} policies in India that has been enforcing comprehensive
affirmative action since 1950. The landmark judgment of the Supreme
Court of India in 2008 mandated that whenever the OBC category (with
27 percent reservation) has unfilled positions, they must be reverted
to general category applicants in admissions to public schools \emph{without
specifying how to implement it}. We disclose the drawbacks of the
recently reformed allocation procedure in admissions to technical
colleges and offer a solution through \emph{``de-reservation via
choice rules}.'' We propose a novel priority design---\emph{Backward
Transfers }(BT) \emph{choice rule}---for institutions and the deferred
acceptance mechanism under these choice rules (DA-BT) for centralized
clearinghouses. We show that DA-BT corrects the shortcomings of existing
mechanisms. By formulating India's legal requirements and policy goals
as formal axioms, we show that the DA-BT mechanism is unique for the
concurrent implementation of reservation and de-reservation policies.
\smallskip{}
\end{abstract}
\begin{elabeling}{00.00.0000}
\item [{$\mathbf{Keywords:}$}] Market design, matching, reserve systems,
de-reservation, affirmative action. 
\end{elabeling}
$\mathbf{JEL\;Codes}$\emph{: }C78, D02, D47, D63, I21.\pagebreak{}

\section{Introduction}

Reserve systems set aside certain available objects/positions for
different \emph{reserve categories}. Each reserve category has its
priority order over individuals. According to a pre-specified order
referred to as a processing sequence, institutions process reserve
categories sequentially to fill their positions.\footnote{Delacrétaz(2021) introduces a model where reserve categories are processed
simultaneously.} Reserve categories allocate their units to the individuals---who
have not yet been allotted a unit---based on their priorities. Priorities
may vary from one reserve category to another to accommodate diversity
concerns. 

Reserve categories are \emph{exclusive} in the sense that only applicants
with certain types or characteristics are considered. It is highly
common for the number of available positions to outnumber the number
of applicants of such exclusive reserve categories. Therefore, objects
may be unassigned, or positions may be unfilled in such exclusive
reserve categories. \emph{$\mathbf{De-reservation}$ }policies have
been introduced along with accompanying reserve policies to alleviate
waste. De-reservation is simply a process of providing unallocated
objects/positions for the use of others. 

De-reservation policies are necessary for many real-world allocation
problems. Moreover, their design and implementation are consequential.
This paper studies the \emph{joint} design and implementation of reserve
systems and de-reservation policies. One of our main contributions
is to bring to light that when the de-reservation policy is not designed
and implemented rigorously, allocation procedures as a whole can have
serious shortcomings. In India's context of admissions to technical
universities, we show how the mechanism currently being used to implement
reserves and de-reservations jointly leads to \emph{sub-optimal outcomes}
concerning applicants' welfare and is \emph{vulnerable to various
types of manipulations}. 

India has been carrying out the most comprehensive affirmative action
program since 1950 to protect historically discriminated sections
of society. The affirmative action program has been implemented via
a reserve system in admissions to publicly funded educational institutions
and allocations of government jobs. Each institution sets aside 15
percent of its slots for applicants from \emph{Scheduled Castes} (SC),
7.5 percent for applicants from \emph{Scheduled Tribes} (ST), and
27 percent for applicants from \emph{Other Backward Classes} (OBC).
These categories are referred to as \emph{reserve }categories\emph{.}
Applicants who do not belong to any reserved category are referred
to as \emph{General Category} (GC). Applicants from reserve categories
must submit their membership information to avail their affirmative
action benefits.\emph{ }Positions earmarked for SC, ST, and OBC are
collectively referred to as \emph{reserve category positions}. The
remaining positions are called \emph{open-category }(or \emph{unreserved})
positions and are available to everyone, including applicants from
reserve categories. Members of SC, ST, and OBC who do not declare
their memberships to their reserve categories and GC members are considered
\emph{only} for open-category positions\textcolor{black}{. }

In the absence of de-reservations, three key principles must be respected
while implementing reservation policy in India:
\begin{enumerate}
\item When applicants from reserve categories obtain open-category positions
these positions are \emph{not} counted against reservations of their
respective categories. This requirement is referred to as the $\mathbf{over-and-above}$
$\mathbf{principle}$. Filling\emph{ }open-category positions before
reserved category positions in each institution achieve this principle. 
\item Each category must respect merit scores in each institution while
allocating their positions. We refer to this requirement as $\mathbf{within-category}$
$\mathbf{fairness}$.\footnote{The Supreme Court of India refers to this requirement as \emph{inter
se merit}.} 
\item Subject to reserve category membership information, each category
must allocate all positions or select all eligible individuals. We
refer to this requirement as $\mathbf{quota-filling}$ $\mathbf{subject}$
$\mathbf{to}$ $\mathbf{eligibility}$. 
\end{enumerate}
In the absence of de-reservations, these requirements identify the
following simple choice procedure: First, applicants are selected
for open-category positions one at a time following the merit score
ranking up to the capacity of open-category. Then, applicants are
selected for SC, ST, and OBC categories following merit score ranking
of applicants in each respective category up to their capacities.
We refer to this choice procedure as the \emph{$\mathbf{India}$ $\mathbf{Reserves}$
$\mathbf{Choice}$ $\mathbf{Rule}$}, or $C^{IN}$ in short. 

In government job allocations and admissions in publicly-funded educational
institutions in India, institutions must respect these legal requirements.
Moreover, in both applications, unfilled SC and ST positions cannot
be provided to others and remain unfilled. However, job allocations
and public school admissions $\mathbf{differ}$ concerning \emph{OBC
de-reservation} policies. In the allocation of government jobs, unfilled
OBC positions remain unfilled. However, the situation changed in admissions
to publicly-funded educational institutions in 2008 with the historic
judgment of the Supreme Court of India (SCI) in\emph{ Ashoka Kumar
Thakur vs. Union of India and} \emph{Others} (2008),\footnote{The judgement is available at https://indiankanoon.org/doc/63489929/.}
which reads: 
\begin{quote}
``\emph{Only non-creamy layer OBCs can avail of reservations in college
admissions, and once they graduate from college they should no longer
be eligible for post-graduate reservation. 27\% is the upper limit
for OBC reservation. The Government need not always provide the maximum
limit. Reasonable cut off marks should be set so that standards of
excellence are not greatly affected. The unfilled seats should revert
to the general category}.'' 
\end{quote}
and, 
\begin{quote}
``\emph{To this end, the Government shall set up a committee to look
into the question of setting the OBC cut off at nor more than 10 marks
below that of the general category. Under such a scheme, whenever
the non-creamy layer OBCs fail to fill the 27\% reservation, the remaining
seats would revert to general category students}.''
\end{quote}
The SCI, however, did \emph{not} specify a well-defined procedure
to revert unfilled OBC positions. There is also a \emph{widespread
ambiguity} regarding potential beneficiaries of reverted OBC positions:
only GC applicants or everyone, including reserve category members.
Lack of specific guidance on dealing with OBC de-reservations leads
to numerous ad-hoc procedures in public school admissions in India.
Technical university seat allocation recently adopted one such ad-hoc
procedure, which we will discuss next. 

\subsection*{2015 Reform in Admissions to Technical Universities}

The admission process at technical universities in India functions
through a centralized marketplace that matches approximately 1.3 million
students to 34,000 university seats. The process was reformed in 2015
with the collaboration of policymakers with a group of computer engineers
and operation researchers. Baswana et al. (2019) discuss the new procedure,
institutional details, and the interaction of the design team with
the policymakers. According to the new assignment procedure, OBC de-reservations
are implemented by\emph{ repeating} the \emph{deferred acceptance
(DA) algorithm} of Gale and Shapley (1962)\emph{ }to adjust capacities
of categories in each program until no OBC seat remains unfilled.
The authors refer to this process as the\emph{ ``Multi-run DA},''
which works as follows:
\begin{itemize}
\item The DA algorithm---in which each program use $C^{IN}$ to select
applicants---is run \emph{on all applicants}. 
\item If there are unfilled OBC seats, then the capacities of reserve categories
are updated by reverting unfilled OBC positions to open-category.
\emph{Re-run the DA algorithm with updated capacitie}s \emph{on all
applicants}.
\item Repeat this procedure until there are no unfilled OBC position in
all programs. 
\end{itemize}
Capacities of open and OBC categories are updated at the end of each
DA run to disentangle OBC de-reservations. The main objective is to
allocate open-category positions (both initially set open-category
positions and reverted surplus OBC positions) before allocating reserve
category positions. This method ensures that the cutoff score for
open-category positions is higher than the cutoff scores of reserve
categories in the presence of de-reservations. We reveal \emph{three
major drawbacks} of the multi-run DA mechanism: 
\begin{enumerate}
\item It causes sub-optimal allocations with respect to applicants' welfare.
\item It is vulnerable to preference manipulation. 
\item It is manipulable by not reporting reserve category membership. 
\end{enumerate}
To fix these shortcomings, we propose a solution through \emph{``de-reservation
via choice rules}''. The idea behind it is to re-run choice rules
to adjust categories' capacities to disentangle de-reservations instead
of re-running the whole mechanism. Specifically, we design a choice
rule---\emph{Backward Transfers (BT) choice rule}--- and the deferred
acceptance mechanism with respect to these choice rules (DA-BT) to
jointly implement reserves and de-reservations. 

The BT choice rule works as follows: Given a set of applicants and
their membership information, a chosen set of applicants is found
via $C^{IN}$. If there is no unfilled OBC seat, the choice procedure
is terminated. If there is a vacancy in the OBC category, capacities
of open-category and OBC are updated by transferring vacancies from
OBC to open-category. Then, $C^{IN}$ is re-run with the updated capacities.
The procedure terminates when there is no unfilled OBC seat. 

There is a simpler \emph{choice rule with reserves} that is outcome
equivalent to the BT choice rule.\footnote{See Aygün and Turhan (2022b) for further discussion.}
Vacant OBC positions can be made available to others by simply modifying
the priority order of OBC positions. However, an important design
criterion is to ensure that open-category's cutoff score is higher
than cutoff scores of reserved categories. When de-reservation is
handled by modifying the priority list of OBC positions, this criterion
is not met.\footnote{See Footnote 10 for further discussion.} 

We show that under the BT choice rule, reserved category applicants
cannot be hurt by reporting their membership to reserve categories
(Theorem 1). The DA-BT mechanism \emph{Pareto dominates} the multi-run
DA mechanism (Theorem 2). Moreover, the DA-BT mechanism is incentive-compatible
(Theorem 3). Therefore, the DA-BT mechanism escapes the shortcomings
of the multi-run DA and presents an unambiguous improvement. We characterize
the DA-BT mechanism with the axioms that echos the legal requirements
and policy perspectives of Indian authorities (Theorem 4). 

The rest of the paper is organized as follows. In Section 2, we present
the model. Section 3 analyzes the multi-run DA mechanism currently
in use and reveals its shortcomings. Section 4 presents the priority
design for institutions, i.e., \emph{the BT choice rule} and the deferred
acceptance mechanism under these choice rules (the DA-BT mechanism).
In the same section, we analyze the properties of the DA-BT mechanism
and provide a characterization. We discuss related literature in Section
5 and, Section 6 concludes. All proofs are presented in the Appendix. 

\section{Framework}

There is a finite set of institutions $\mathcal{S}=\left\{ s_{1},...,s_{m}\right\} $
and a finite set of individuals $\mathcal{I}=\left\{ i_{1},...,i_{n}\right\} $.
Institution $s$ has $\overline{q}_{s}$ positions. The vector $\left(q_{s}^{SC},q_{s}^{ST},q_{s}^{OBC}\right)$
denotes the number of positions earmarked for SC, ST, and OBC categories
at institution $s$. We let $\mathcal{R}=\left\{ SC,ST,OBC\right\} $
to denote the set of reserve categories, and let $\mathcal{C}=\left\{ o,SC,ST,OBC\right\} $
to denote the set of all position categories, where $o$ denotes the
open-category.\footnote{Our model can be generalized to to a model with an arbitrary number
of reserve categories among which an arbitrary number of them can
be de-reserved. Our results are independent of the number of reserve
categories considered and hold in a more general model. We choose
to formulate the admissions to engineering colleges in India as it
is the main focus of this manuscript. } The number of open-category positions at institution $s$ is $q_{s}^{o}=\overline{q}_{s}-q_{s}^{SC}-q_{s}^{ST}-q_{s}^{OBC}$.
The vector $q_{s}=\left(q_{s}^{o},q_{s}^{SC},q_{s}^{ST},q_{s}^{OBC}\right)$
describes the \emph{initial} \emph{distribution} of positions over
reserve categories at institution $s$. Let $\mathbf{q}=\left(q_{s}\right)_{s\in\mathcal{S}}$
denote a profile of vectors for the initial distribution of positions
over categories at institutions.\footnote{There is also Supreme Court mandated reservations for persons with
indivisibilities that are implemented within each reserved category,
including the open-category. For simplicity, we do not model reservations
for persons with indivisibilities in this paper. The focus of this
paper is the joint implementation of caste-based reservations and
OBC de-reservations. However, our model can straightforwardly be extended
to a model that includes reservations for persons with indivisibilities. } 

The function $t:\mathcal{I}\rightarrow\mathcal{\mathcal{R}}\cup\left\{ GC\right\} $
denotes the category membership of individuals. For every individual
$i\in\mathcal{I}$, $t(i)$, or $t_{i}$, denotes the category individual
$i$ belongs to. Reporting membership to SC, ST, and OBC is $\mathbf{optional}$.
Reserved category members who do not report their membership are considered
GC applicants and eligible $\mathbf{only}$ for open-category positions.
Members of reserve category $r\in\mathcal{R}$ are eligible for $\mathbf{both}$
open-category positions and reserved category $r$ positions. We denote
a profile of reserved category membership by $T=\left(t_{i}\right)_{i\in\mathcal{I}}$,
and let $\mathcal{T}$ be the set of all possible reserved category
membership profiles. 

Merit scores induce strict\emph{ }meritorious ranking of individuals
at each institution $s$, denoted $\succ_{s}$, which is a linear
order over $\mathcal{I}\cup\{\emptyset\}$. $i\succ_{s}j$ means that
applicant $i$ has a higher priority (higher merit score) than applicant
$j$ at institution $s$. We write $i\succ_{s}\emptyset$ to say that
applicant $i$ is acceptable for institution $s$. Similarly, we write
$\emptyset\succ_{s}i$ to say that applicant $i$ is unacceptable
for institution $s$. The profile of institutions' priorities is denoted
$\succ=(\succ_{s_{1}},...,\succ_{s_{m}})$. 

For each institution $s\in\mathcal{S}$, the merit ordering for individuals
of type $r\in\mathcal{R}$, denoted by $\succ_{s}^{r}$, is obtained
from $\succ_{s}$ in a straightforward manner as follows: 
\begin{itemize}
\item for $i,j\in\mathcal{I}$ such that $t_{i}=r$, $t_{j}\neq r$, $i\succ_{s}\emptyset$,
and $j\succ_{s}\emptyset$, we have $i\succ_{s}^{r}\emptyset\succ_{s}^{r}j$,
where $\emptyset\succ_{t}^{r}j$ means individual $j$ is unacceptable
for category $r$ at institution $s$. 
\item for any other $i,j\in\mathcal{I}$, $i\succ_{s}^{r}j$ if and only
if $i\succ_{s}j$. 
\end{itemize}
Each individual $i\in\mathcal{I}$ has a strict preference relation
$P_{i}$ over $\mathcal{S}\cup\left\{ \emptyset\right\} $, where
$\emptyset$ denotes remaining unmatched. We write $sP_{i}\emptyset$
to mean that institution $s$ is \emph{acceptable }for individual
$i$. Similarly, $\emptyset P_{i}s$ means institution $s$ is \emph{unacceptable
}for individual $i$. We denote the profile of true individual preferences
by $P=\left(P_{i}\right)_{i\in\mathcal{I}}$. We denote by $R_{i}$
the weak preference relation associated with the strict preference
relation $P_{i}$, and by $R=\left(R_{i}\right)_{i\in\mathcal{I}}$
the profile weak preferences. Note that we assume that reserve category
individuals only care about their matched institution, and are indifferent
about open-category and reserve category positions. 

For each institution $s\in\mathcal{S}$, its selection criterion is
summarized by a choice rule $C_{s}$. A choice rule $C_{s}$ simply
selects a subset from any given set of individuals. That is, for any
given $A\subseteq\mathcal{I}$, $C_{s}\left(A\right)\subseteq A$. 

We now define three axioms that are legally required principles for\emph{
resource allocation problems in India in the absence of de-reservations}.
These axioms are concerned with institutions' choice functions. Given
a set of applicants $A\subseteq\mathcal{I}$, let $rank_{A}(i)$ be
the rank of applicant $i$ in set $A$ with respect to merit-ranking
$\succ_{s}$. That is, $rank_{A}(i)=k$ if and only if $\mid\left\{ j\in A\mid j\succ_{s}i\right\} \mid=k-1$.
Let $A^{r}=\left\{ i\in A\mid i\succ_{s}^{r}\emptyset\right\} $ be
the set of category $r\in\mathcal{R}$ eligible individuals in set
$A$. 

\paragraph{Over-and-above principle. }

Each individual $i\in A$ with $rank_{A}(i)\leq q_{s}^{o}$ must be
assigned to an open-category position. 

This property ensures that open-category positions obtained by reserve
category members are not counted against their reservations. By allocating
open-category positions to highest-scoring applicants, reserve category
positions can be given to members of reserve categories who would
not be able to receive these positions in the absence of reservation
policy. Moreover, over-and-above principle guarantees that cutoff
score of open-category is higher than cutoff scores of reserve categories. 

\paragraph{Within-category fairness.}

Given two individuals $i,j\in A$ such that $t_{i}=t_{j}$ and $i\succ_{s}j$,
if $j$ is assigned a position, then $i$ must also be assigned a
position. 

Within-category fairness requires that merit scores of applicants
are respected in each reserve category. That is, if a lower-scoring
applicant receives a position from category $c$, then a higher-scoring
applicant with the same category membership must receive a position
as well. 

The last axiom is a desirable efficiency requirement. 

\paragraph{Quota-filling subject to eligibility. }

If individual $i$ with $t_{i}=r$ is unassigned, then the number
of individuals who are matched to a category $r$ position must be
equal to $q_{s}^{r}$, for all $r\in\mathcal{R}$. 

In the absence of de-reservations, these legal requirements uniquely
induce the following choice rule:

\subsubsection*{India Reserves Choice Rule $C_{s}^{IN}$}

Given an initial distribution of positions $q_{s}=\left(q_{s}^{o},q_{s}^{SC},q_{s}^{ST},q_{s}^{OBC}\right)$,
a set of applicants $A\subseteq\mathcal{I}$, and a profile reserve
category membership $T=\left(t_{i}\right)_{i\in A}$ for the members
of $A$, the set of chosen applicants $C_{s}^{IN}(A,q_{s})$, is computed
as follows:

\paragraph{Stage 1. }

For open-category positions, individuals are selected following $\succ_{s}$
up to the capacity $q_{s}^{o}$. Let $C_{s}^{o}\left(A,q_{s}^{o}\right)$
be the set of chosen applicants. 

\paragraph{Stage 2. }

Among the remaining applicants $A^{'}=A\setminus C_{s}^{o}\left(A,q_{s}^{o}\right)$,
for each reserve category $t\in\mathcal{R}$, applicants are chosen
following $\succ_{s}^{t}$ up to the capacity $q_{s}^{t}$. Let $C_{s}^{t}\left(A^{'},q_{s}^{t}\right)$
be the set of chosen applicants for reserve category $t$.

Then, $C_{s}^{IN}(A,q_{s})$ is defined as the union of the set of
applicants chosen in stages 1 and 2. That is, 
\[
C_{s}^{IN}(A,q_{s})=C_{s}^{o}\left(A,q_{s}^{o}\right)\cup\underset{t\in\mathcal{R}}{\bigcup}C_{s}^{t}\left(A^{'},q_{s}^{t}\right).
\]

\subsection{Matching, Assignment, and Mechanisms}

A choice rule determines who is chosen from any given set of individuals
when there is a single institution. Admissions to technical universities
is a centralized marketplace with multiple institutions. An outcome
in a centralized marketplace is a matching. 

\paragraph{Matching. }

A $\mathbf{matching}$ $\mu$ is a function $\mu:\mathcal{I}\cup\mathcal{S}\rightarrow2^{\mathcal{I}}\cup\mathcal{S}\cup\left\{ \oslash\right\} $
such that 
\begin{enumerate}
\item for any individual $i\in\mathcal{I}$, $\mu_{i}\in\mathcal{S}\cup\{\oslash\}$,
\item for any institution $s\in\mathcal{S}$, $\mu_{s}\in2^{\mathcal{I}}$
such that $\mid\mu_{s}\mid\leq\overline{q}_{s}$, 
\item for any individual $i\in\mathcal{I}$ and institution $s\in\mathcal{S}$,
$\mu_{i}=s$ if and only if $i\in\mu_{s}$. 
\end{enumerate}
A matching specifies, for every institution, the set of individuals
who are assigned to that institution. A matching does \emph{not} specify
categories under which individuals are assigned. 

\paragraph{Assignment. }

Associated with a matching is an \emph{$\mathbf{assignment}$} which
specifies a category each individual is accepted under in each institution.\emph{
}Each individual's assignment is a $\mathbf{pair}$ of institution
and category, and each institution's assignment is a set of individual-category
pairs. 

Formally, an assignment is a function $\eta:\;\mathcal{I}\cup\mathcal{S}\rightarrow\left(2^{\mathcal{I}}\cup\mathcal{S}\right)\times\mathcal{C}\bigcup\left\{ \oslash\right\} $
such that 
\begin{enumerate}
\item for any $i\in\mathcal{I}$, 
\[
\begin{cases}
\begin{array}{c}
\eta(i)\in\mathcal{S}\times\left\{ o\right\} \bigcup\left\{ \oslash\right\} \\
\eta(i)\in\mathcal{S}\times\left\{ t(i),o\right\} \bigcup\left\{ \oslash\right\} 
\end{array} & \begin{array}{c}
if\;t_{i}=GC,\\
if\;t_{i}\in\mathcal{R},
\end{array}\end{cases}
\]
\item for any $s\in\mathcal{S}$, $\eta(s)\subseteq2^{\mathcal{I}\times\mathcal{C}}$
such that $\mid\eta(s)\mid\leq\overline{q}_{s}$, and for all $r\in\mathcal{R}$,
\[
\mid\left\{ j\mid\left(j,r\right)\in\eta(s)\right\} \mid\leq q_{s}^{r},
\]
 
\item for every individual $i\in\mathcal{I}$ and institution $s\in\mathcal{S}$,
$\eta(i)=\left(s,c\right)$ if and only if $\left(i,c\right)\in\eta(i)$. 
\end{enumerate}
Let $\mu\left(\eta\right)$ be the matching induced by the assignment
$\eta$ and $\mu_{i}\left(\eta\right)$ be the institution that individual
$i$ is matched with. Similarly, $\mu_{s}\left(\eta\right)$ denotes
the set of individuals who are matched with institution $s$. Given
an assignment $\eta$, the matching $\mu\left(\eta\right)$ induced
by it is obtained as follows: 
\begin{itemize}
\item $\mu_{i}\left(\eta\right)=s$ if and only if $\eta(i)=\left(s,c\right)$
for some $c\in\mathcal{C}$, and 
\item $\mu_{s}(\eta)=\left\{ i\in\mathcal{I}\mid\left(i,c\right)\in\eta\left(s\right)\quad for\;some\quad c\in\mathcal{C}\right\} $.
\end{itemize}

\paragraph{Mechanisms. }

A \emph{$\mathbf{mechanism}$ }is a systematic way to map preference
and reserve category membership profiles of individuals to assignments,
given institutions' choice procedures. Technically, a mechanism $\varphi$
is a function $\varphi:\mathcal{P}\times\mathcal{T}\rightarrow\mathcal{M}$,
where $\mathcal{M}$ denotes the set of all assignments, given a profile
of institutional choice rules $\mathbf{C}=(C^{s})_{s\in\mathcal{S}}$.
Note that the outcome of a mechanism is an assignment, not a matching.
In India, outcomes are announced as institution-category pairs for
individuals. 

A mechanism $\mathcal{\varphi}$ is \emph{$\mathbf{incentive-compatible}$
}if for every profile $(P,T)\in\mathcal{P}\times\mathcal{T}$, and
for each individual $i\in\mathcal{I}$, there is no $\left(\widetilde{P}_{i},\widetilde{t}_{i}\right)$,
such that 
\[
\mu_{i}\left[\varphi\left(\left(\widetilde{P}_{i},\widetilde{t}_{i}\right),\left(P_{-i},T_{-i}\right)\right)\right]P_{i}\mu_{i}\left[\varphi\left(P,T\right)\right].
\]

That is, a mechanism is incentive-compatible if reporting the\emph{
true preference} and \emph{true reserve category membership} pair
is a weakly dominant strategy for each individual. Affirmative action
policies are designed to increase admission chances of reserve category
members in the sense that the assignment of a reserve category member
when she claims her membership is at least as good as the assignment
she would receive without reporting her membership. That is, reporting
their membership to reserve categories should not hurt them. Otherwise,
the rationale behind the affirmative action policy is violated. 

We now describe the DA algorithm with respect to India Reserves choice
rules, which will prove useful for describing the multi-run DA mechanism. 

\subsubsection*{DA Mechanism with India Reserves Choice Rules}

Suppose that $P=\left(P_{i}\right)_{i\in\mathcal{I}}$ is the vector
of the\emph{ reported} preference relations and $T=\left(t_{i}\right)_{i\in\mathcal{I}}$
is a  vector of reported reserve category membership of individuals.
Given institutions' priority rankings $\succ=\left(\succ_{s}\right)_{s\in\mathcal{S}}$
and the profile $\mathbf{q}=\left(q_{s}\right)_{s\in\mathcal{S}}$---therefore,
given the India reserves choice function of each institution $s\in\mathcal{S}$,
$C_{s}^{IN}$---the outcome of the DA mechanism with India Reserves
rules is computed as follows: 

\paragraph{Step 1. }

Each individual in $\mathcal{I}$ applies to his top choice institution.
Let $\mathcal{A}_{s}^{1}$ be the set of individuals who apply to
institution $s$, for each $s\in\mathcal{S}$. Each institution $s$
holds  applicants in $C_{s}^{IN}\left(\mathcal{A}_{s}^{1},q_{s}\right)$
and rejects the rest. 

\paragraph{Step n$\protect\geq2$.}

Each individual who was rejected in the previous step applies to the
best institution that has not rejected him. Let $\mathcal{A}_{s}^{n}$
be the union of the set of individuals who were tentatively held by
institution $s$ at the end of Step $n-1$ and the set of new proposers
of $s$ in Step $n$. Each institution $s\in\mathcal{S}$ tentatively
holds individuals in $C_{s}^{IN}\left(\mathcal{A}_{s}^{n},q_{s}\right)$
and rejects the rest. 

The DA algorithm terminates when there are no rejections. The outcome
is the tentative assignments at that point. We denote the outcome
by $\Phi(P,\mathbf{q})$ to emphasize the dependence of the outcome
on the vector of institutional reserve structures given by the profile
$\mathbf{q}$. We denote $\Phi_{i}\left(P,\mathbf{q}\right)$ be the
assignment of individual $i$. 

\section{Implementation of Reservations and De-reservations }

In this section we discuss the SCI's judgement in Ashoka Kumar Thakur
(2008) and its particular implementation in admissions to technical
universities which was reformed in 2015. 

\subsection{Impossibility Implied by Ashoka Kumar Thakur (2008) }

In Ashoka Kumar Thakur (2008), the SCI ruled that the remaining seats
would revert to \emph{general category} students whenever the OBCs
fail to fill the 27\% reservation. Even though it aims to reduce waste
in instances of low demand from OBC category applicants, there are
unintended consequences, as is stated. Before explaining the impossibility
induced by Ashoka Kumar Thakur (2008), we first define a crucial criterion
for implementing affirmative action programs. 
\begin{defn}
A choice rule $C$ $\mathbf{complies}$ $\mathbf{with}$ $\mathbf{the}$
$\mathbf{spirit}$ $\mathbf{of}$ $\mathbf{affirmative}$ $\mathbf{action}$
if, for any given set of individuals $A\subseteq\mathcal{I}$ and
any member $i$ of reserve category $r\in\mathcal{R}$,  if $i\notin C(A)$
when $i$ reports her membership to category $r$, then $i\notin C(A)$
when $i$ does not report her membership to $r$. 
\end{defn}
A selection criterion complies with the spirit of affirmative action
if reporting their membership to reserve categories should not hurt
reserve category members. Otherwise, members of reserve categories
may have an incentive not to state their category memberships and
avail of affirmative action concessions. This property is a simpler
version of the ''privilege monotonicity'' introduced in Aygün and
Bó (2021). 

No choice rule satisfies the over-and-above principle, within-category
fairness, quota-filling subject to eligibility, complies with the
spirit of affirmative action, and implements the Ashoka Kumar Thakur's
(2008) decision as stated. Given the legal requirements, there is
an incompatibility between the SCI's decree and the property of complying
with the spirit of affirmative action. To see it consider the following
simple example:
\begin{example}
Institution $s$ has six positions with the distribution of positions
$\left(q_{s}^{o},q_{s}^{SC},q_{s}^{OBC}\right)=\left(3,1,2\right).$
There are seven applicants with following category memberships and
exam scores: 

\[
\begin{array}{ccc}
Applicant & Category & Score\\
i_{1} & GC & 99\\
i_{2} & GC & 98\\
i_{3} & GC & 97\\
i_{4} & OBC & 96\\
i_{5} & SC & 95\\
i_{6} & SC & 94\\
i_{7} & GC & 93
\end{array}
\]

The over-and-above principle, within-category fairness, and quota-filling
subject to eligibility imply that open-category positions are assigned
to $\left\{ i_{1},i_{2},i_{3}\right\} $, $i_{4}$ is assigned to
one of reserved OBC positions, and $i_{5}$ is assigned to reserved
SC position. One of the reserved OBC positions remains unfilled. Following
Ashoka Kumar Thakur (2008) it must be given to applicant $i_{7}$.
In this case, applicant $i_{6}$ has an incentive not to report her
SC membership and to be considered as a GC applicant. When $i_{6}$
participates as a GC candidate, she will be assigned to the unfilled
OBC position and causes $i_{7}$ to be unassigned. There is no choice
procedure that complies with the spirit of affirmative action when
leftover positions are to be given to GC applicants. 
\end{example}
The straightforward fix is to allocate surplus OBC positions as \emph{open-category
positions} on the basis of merit. Indeed, in admissions to technical
universities unfilled OBC positions are allocated as open-category
positions following the merit scores. The following quote from the
''\emph{Business Rules for Joint Seat Allocation for the Academic
Programs offered by the IITs, NITs, IIESTs, IIITs, and Other-GFTIs
for the academic year 2021-22}''\footnote{The report can be accessed at https://josaa.nic.in/webinfo/File/GetFile//?FileId=1\&LangId=P
(last accessed on 12-16-2021).} clearly indicates that de-reserved positions are allocated as open-category
positions which are available for reserve category members as well: 
\begin{quote}
\emph{``Unfilled GEN-EWS and OBC-NCL category seats will be de-reserved
and treated as OPEN category seats for allocation in every round of
seat allocation. However, unfilled SC and ST category seats will NOT
be de- reserved}''.
\end{quote}

\subsection{Implementation of Reservations and De-reservations via Multi-run
DA }

Baswana et al. (2018) report the new design for the joint seat allocation
process for the technical universities in India that has been implemented
since 2015. A sequential procedure---called \emph{multi-run DA}---is
introduced to deal with de-reservations. According to multi-run DA,
the DA algorithm is first run with the initial capacities of reserve
categories at each program. If there are unfilled seats that can be
de-reserved in the resulting matching, then the capacities are updated
by transferring the unfilled seats to open-category. Then, the DA
is \emph{re-run} on all individuals with \emph{updated} capacities
of reserve categories at each institution. If there are no vacant
seats that can be de-reserved in the resulting matching, then the
process is terminated. 

We now formulate the multi-run DA algorithm. Let $\mathbf{q}=\mathbf{q}^{1}$
be the profile of initial distribution vector of positions over categories.
Given a vector of reported preference relations of applicants $P=\left(P_{i}\right)_{i\in\mathcal{I}}$,
a vector of reported reserve category membership $T=\left(t_{i}\right)_{i\in\mathcal{I}}$,
and a profile of institutions' choice rules $\left(C_{s}^{IN}\right)_{s\in\mathcal{S}}$,
the multi-run DA algorithm runs as follows:

\paragraph{Iteration 1. }

Run the DA with initial distributions of positions over categories
$\mathbf{q}^{1}$. Each institution $s\in\mathcal{S}$, use $C_{s}^{IN}\left(\cdot,q_{s}^{1}\right)$
to select applicants. Let $r_{s}^{1}$ be the number of vacant OBC
seats. Update the number of open-category and OBC positions by transferring
$r_{s}^{1}$ many positions from OBC to open-category. Let $\mathbf{q}^{2}$
be the profile of updated distributions of positions over categories. 

\paragraph{Iteration n (n$\protect\geq$2).}

Run the DA with the updated distributions of reserved categories $\mathbf{q}^{n}$.
Each institution $s\in\mathcal{S}$ use $C_{s}^{IN}\left(\cdot,q_{s}^{n}\right)$
to select applicants. Let $r_{s}^{n}$ be the number of vacant OBC
seats. Update the number of open-category and OBC positions by transferring
$r_{s}^{n}$ many positions from OBC to open-category. Let $\mathbf{q}^{n+1}$
be the profile of updated distributions of positions over categories. 

The algorithm terminates when there is no vacant position that can
be de-reserved at any institution. We denote the outcome of multi-run
DA by $\Phi\left(P,\mathbf{q}^{L}\right)$, where $L$ is the number
of iterations needed, and $\mathbf{q}^{L}$ denotes the profile of
updated distribution of positions at institutions in the last iteration.
The outcome of individual $i\in\mathcal{I}$ is denoted by $\Phi_{i}\left(P,\mathbf{q}^{L}\right)$. 

To demonstrate how the multi-run DA algorithm is run, we present the
following example. 
\begin{example}
There are two institutions $a$ and $b$, both of which have two positions.
Each institution reserves one position for OBC and consider the other
as open-category. There are four applicants: $i_{1}$, $i_{2}$, $i_{3}$,
and $i_{4}$. Suppose $t_{i_{1}}=t_{i_{2}}=GC$ and $t_{i_{3}}=t_{i_{4}}=OBC$.
The merit scores of applicants are ranked from highest to lowest as
$i_{1}-i_{2}-i_{3}-i_{4}$. Preferences of applicants are given in
the table below. 
\[
\begin{array}{cccc}
i_{1} & i_{2} & i_{3} & i_{4}\\
a & a & b & b\\
b & \emptyset & a & a\\
\emptyset &  & \emptyset & \emptyset
\end{array}
\]
 
\end{example}

\paragraph{\emph{Iteration 1: }}

Individuals $i_{1}$ and $i_{2}$ apply to institution $a$, while
$i_{3}$ and $i_{4}$ apply to institution $b$. In institution $a$,
$i_{1}$ is tentatively held for the open-category position and $i_{2}$
is rejected because she is not eligible for OBC position. In institution
$b$, $i_{3}$ is tentatively held for the open-category position,
and $i_{4}$ is tentatively held for the OBC position. Since $i_{2}$
does not find any other institution as acceptable, the DA outcome
in the first iteration is $\left(\begin{array}{cc}
a & b\\
\left\{ i_{1}\right\}  & \left\{ i_{3},i_{4}\right\} 
\end{array}\right)$. Since there is a vacant OBC slot in institution $a$, it is reverted
to open-category. Hence, distribution of positions in $a$ is updated
from $(1,1)$ to $(2,0)$, while distribution of positions in $b$
remains $(1,1)$. 

\paragraph{Iteration 2: }

Both $i_{1}$ and $i_{2}$ are tentatively held by the open-category
positions in $a$. Applicants $i_{3}$ and $i_{4}$ are held by $b$
in open-category and OBC, respectively. Hence, the outcome of the
second run is $\left(\begin{array}{cc}
a & b\\
\left\{ i_{1},i_{2}\right\}  & \left\{ i_{3},i_{4}\right\} 
\end{array}\right)$. Each applicant is assigned to her top choice institution. 

\subsection*{Case Against the Multi-run DA}

We disclose three shortcomings of the multi-run DA: (1) inefficiency,
(2) manipulability via preference misreporting, and (3) manipulability
via not reporting reserve category membership. 

\subsubsection*{Inefficiency}

Consider the following example to see the inefficiency of multi-run
DA. 
\begin{example}
There are two institutions $a$ and $b$, each of which has two positions.
Both reserve one position for OBC and consider the other as open-category.
There are four applicants: $i_{1}$, $i_{2}$, $i_{3}$, and $i_{4}$.
Suppose $t_{i_{1}}=t_{i_{2}}=GC$ and $t_{i_{3}}=t_{i_{4}}=OBC$.
The merit scores of applicants are ranked from highest to lowest as
$i_{1}-i_{2}-i_{3}-i_{4}$. Applicants' true preferences are given
below: 
\[
\begin{array}{cccc}
i_{1} & i_{2} & i_{3} & i_{4}\\
a & a & b & b\\
b & b & a & a\\
\emptyset & \emptyset & \emptyset & \emptyset
\end{array}
\]
 In the first iteration of DA, applicants $i_{1}$ and $i_{2}$ are
considered in institution $a$, while applicants $i_{3}$ and $i_{4}$
are considered in institution $b$. Since $i_{1}$ and $i_{2}$ are
GC candidates, they are considered only for an open-category seat
in institution $a$. $i_{1}$ is tentatively held for the open-category
seat while $i_{2}$ is rejected. In institution $b$, applicant $i_{3}$
is tentatively held by the open-category seat and applicant $i_{4}$
is tentatively held by the OBC seat. Now, $i_{2}$ applies to $b$.
Institution $b$ holds $i_{2}$ for the open-category seat and $i_{3}$
for the OBC seat. Applicant $i_{4}$ is rejected from $b$ in return.
Next, $i_{4}$ applies $a$ and is held by the OBC seat. The outcome
is $\left(\begin{array}{cc}
a & b\\
\left\{ i_{1},i_{4}\right\}  & \left\{ i_{2},i_{3}\right\} 
\end{array}\right)$. The first iteration of the deferred acceptance is the final iteration
since no de-reservation occurs. 
\end{example}
Applicants $i_{2}$ and $i_{4}$ are assigned their least favorite
institutions. It is caused by an \emph{unnecessary rejection chain},
which can be prevented by carefully executing de-reservations within
institutions' choice rules.\footnote{Kesten's (2010) \emph{Efficiency Adjusted Deferred Acceptance }algorithm
provides efficiency gain over student-proposing deferred acceptance
algorithm by preventing unnecessary rejection chains via students'
consent decisions. Efficiency gain over the multi-run DA is obtained
by re-designing institutions' selection procedures, which we explain
in Section 4. }

\subsubsection*{Manipulability via preference misreporting}

Multi-run DA is vulnerable to manipulation via preference misreporting.
Consider Example 4. Note that applicant $i_{2}$ is assigned to $b$,
i.e., her second choice institution. Now, consider the following preferences,
where $i_{2}$ misreports by stating $a$ as the only acceptable alternative.
That is, applicants' stated preferences are as follows: 
\[
\begin{array}{cccc}
i_{1} & i_{2} & i_{3} & i_{4}\\
a & a & b & b\\
b & \emptyset & a & a\\
\emptyset &  & \emptyset & \emptyset
\end{array}
\]
 This is the same market as in Example 3. The outcome of multi-run
DA is $\left(\begin{array}{cc}
a & b\\
\left\{ i_{1},i_{2}\right\}  & \left\{ i_{3},i_{4}\right\} 
\end{array}\right)$, where each applicant is assigned their top choices. Therefore, by
misreporting, applicant $i_{2}$ receives a strictly better outcome. 

\subsubsection*{Manipulability via not revealing reserve category membership}

The multi-run DA mechanism provides an advantage to individuals who
can strategize by not revealing their reserve category membership\emph{.
}To see it consider Example 3, where the multi-run DA outcome is $\left(\begin{array}{cc}
a & b\\
\left\{ i_{1},i_{4}\right\}  & \left\{ i_{2},i_{3}\right\} 
\end{array}\right)$ when both $i_{3}$ and $i_{4}$ truthfully report their OBC membership
under the given true preference profile. Now suppose that individual
$i_{4}$ does not report her OBC membership, and, therefore, she is
considered only for open-category positions. 

In the first iteration of DA, individuals $i_{1}$ and $i_{2}$ apply
to institution $a$, while applicants $i_{3}$ and $i_{4}$ apply
to institution $b$ in the first step. Since $i_{1}$ and $i_{2}$
are GC candidates, they are considered only for an open-category seat
in institution $a$. $i_{1}$ is tentatively held for the open-category
seat while $i_{2}$ is rejected. In institution $b$, both $i_{3}$
and $i_{4}$ are first considered for the open-category position.
Since $i_{3}$ has a higher score, $i_{4}$ gets rejected. Note that
since $i_{4}$ did not claim her OBC membership, she gets rejected
from institution $b$. In the second step of the DA, $i_{2}$ applies
to $b$ and $i_{4}$ applies to $a$. At institution $b$, individual
$i_{2}$ receives the open-category position by replacing $i_{3}$
and $i_{3}$ receives the reserved OBC slot. At institution $a$,
$i_{1}$ keeps her open-category position and $i_{4}$ is rejected.
Therefore, the first DA iteration results in the outcome $\left(\begin{array}{cc}
a & b\\
\left\{ i_{1}\right\}  & \left\{ i_{2},i_{3}\right\} 
\end{array}\right)$. Since the OBC position in institution $a$ remains unfilled, it
is set as an open-category position for the second iteration of DA. 

We now run the second DA iteration on all individuals. $i_{1}$ and
$i_{2}$ apply to $a$, and they are both assigned to open-category
positions since $a$ has two open-category positions now. $i_{3}$
and $i_{4}$ apply to institution $b$. $i_{3}$ is assigned to the
open-category position and $i_{4}$ gets rejected since she can be
considered only for open-category positions and has a lower score
than $i_{3}$. In the second step of the second iteration of DA, $i_{4}$
applies to her second choice, i.e., institution $b$. However, she
gets rejected because her score is lower than both $i_{1}$ and $i_{2}$.
Therefore, the second iteration DA outcome is $\left(\begin{array}{cc}
a & b\\
\left\{ i_{1},i_{2}\right\}  & \left\{ i_{3}\right\} 
\end{array}\right)$. Since the OBC seat remains vacant in $b$, it is set to an open-category
position so that $b$ now has two open-category positions. 

In the third iteration of DA, both $a$ and $b$ have two open-category
positions. $i_{1}$ and $i_{2}$ apply to these and they are both
assigned to open-category positions. $i_{3}$ and $i_{4}$ apply to
$b$ and they are both assigned to open-category positions. Therefore,
the outcome of the third DA iteration is $\left(\begin{array}{cc}
a & b\\
\left\{ i_{1},i_{2}\right\}  & \left\{ i_{3},i_{4}\right\} 
\end{array}\right)$.

Note that when $i_{4}$ truthfully reveals her OBC category membership
she was assigned to institution $a$, which is her second choice.
However, when she does not reveal her OBC category membership she
is assigned to her top choice, institution $b$. 

The purpose of the reservation policy is to protect the members of
SC, ST, and OBC communities when they claim their privilege. This
example, however, shows that it is possible for a reserved category
member to get assigned to a better institution by not claiming her
affirmative action privilege. 

\section{De-reservation via Choice Rules}

We present a priority design to jointly implement reservation and
de-reservation policies in a single institution. The multi-run DA
mechanism presents policymakers\textquoteright{} perspective that
reverted OBC positions should be allocated as open-category positions
before reserve category positions. According to this perspective,
reserve category members can benefit from these extra positions, and
positions taken by reserve category members are not counted against
their categories\textquoteright{} reservations. Moreover, it ensures
that, in the presence of de-reservations, cutoff score of open-category
is higher than the cutoff scores of reserve categories. We argue that
disentangling de-reservations within institutions\textquoteright{}
choice rules by re-running choice rules improves applicants\textquoteright{}
welfare and removes manipulation possibilities via preference misreporting
or not reporting reserve category membership.

We now present the \emph{Backward Transfers }choice rule, denoted
$C^{BT}$ for a single institution. 

\subsection*{Backward Transfers Choice Rule $C_{s}^{BT}$}

Consider institution $s$. Let $A\subseteq\mathcal{I}$ be a set of
applicants. Given a vector of reported reserve category membership
$T=\left(t_{i}\right)_{i\in A}$, and a vector of initial distribution
of positions over reserve categories $q_{s}=q_{s}^{1}$, $C_{s}^{BT}$
selects applicants in multiple iterations as follows: 

\paragraph{Iteration 1.}

Run $C_{s}^{IN}\left(A,q_{s}^{1}\right)$. Let $\tau^{1}$ be the
number of vacant OBC positions. If $\tau^{1}=0$, then the procedure
terminates and $C_{s}^{IN}\left(A,q_{s}^{1}\right)$ is the set of
chosen applicants. If $\tau^{1}\geq1$, then we update the number
of open-category and OBC positions by transferring $\tau^{1}$ many
positions from OBC to open-category. Let $q_{s}^{2}$ be the vector
of updated distributions of positions over reserve categories. 

\paragraph{Iteration n (n$\protect\geq$2).}

Run $C_{s}^{IN}\left(A,q_{s}^{n}\right)$, where $q_{s}^{n}$ is the
updated distribution of positions over reserve categories. Let $\tau^{n}$
be the number of vacant OBC seats. If $\tau^{n}=0$, then the procedure
terminates and $C_{s}^{IN}\left(A,q_{s}^{n}\right)$ is the set of
chosen applicants. If $\tau^{n}\geq1$, then we update the number
of open-category and OBC positions by transferring $\tau^{n}$ positions
from OBC to open-category. Let $q_{s}^{n+1}$ be the vector of updated
distributions of positions over reserve categories. 

The choice process terminates when there is no vacant OBC position
at any institution. The set of applicants who are selected in the
last iteration---call it $N$---are the applicants who are selected
by the backward transfers choice rule.\footnote{One should notice that an outcome equivalent choice procedure to the
Backward Transfers Choice Rule can be defined by modifying the priorities
of OBC slots. This possibility was offered to Indian authorities by
Baswana et al. (2018), but ultimately rejected. Authorities' main
concern was to keep open-category cutoff scores higher than OBC cutoff
scores. Modifying OBC priorities could violate the desired ranking
in  cutoff scores.\emph{ }Baswana et al. (2018) report their interaction
with the Indian policy makers as follows: 
\begin{quote}
``\emph{Business rule 5 required unfilled OBC seats to be made available
to Open category candidates. The approach we initially suggested involved
construction of augmented Merit Lists making Open category candidates
eligible for OBC seats but at a lower priority than all OBC candidates,
and modification of virtual preference lists so that general candidates
now apply for both the OPEN and the OBC virtual programs. We showed
that running our algorithm on these modified inputs would produce
the candidate optimal allocation satisfying the business rules. However,
}\textcolor{black}{\emph{the authorities feared that this approach
may cause issues}}\emph{ with computing the closing rank correctly
(see Design Insight 6), or have some other hidden problem. An authority
running centralized college or school admissions is typically loathe
to modify, add complexity to, or replace software that is tried and
tested}.'' 
\end{quote}
} That is,

\[
C_{s}^{BT}(A,q_{s})=C_{s}^{IN}(A,q_{s}^{N}).
\]

To demonstrate how the backward transfers choice rule is run, consider
the following example.
\begin{example}
Consider institution $s$ that has eight positions with the distribution
of positions 
\[
q_{s}=\left(q_{s}^{o},q_{s}^{SC},q_{s}^{ST},q_{s}^{OBC}\right)=\left(3,1,1,3\right).
\]
 Let $A=\left\{ i_{1},i_{2},i_{3},i_{4},i_{5},i_{6},i_{7},i_{8}\right\} $
be the set of individuals with following category memberships and
exam scores: 

\[
\begin{array}{ccc}
Individual & Category & Score\\
i_{1} & GC & 100\\
i_{2} & GC & 99\\
i_{3} & GC & 98\\
i_{4} & OBC & 97\\
i_{5} & SC & 96\\
i_{6} & ST & 95\\
i_{7} & GC & 94\\
i_{8} & GC & 93
\end{array}
\]

The outcome of $C_{s}^{BT}$ is computed as follows: 
\end{example}

\paragraph{\emph{Iteration 1}: }

$\{i_{1},i_{2},i_{3}\}$ are assigned open-category positions. $i_{4}$
is assigned to one of the reserved positions for OBC. \emph{Two reserved
OBC positions remain unfilled}. That is, $\tau^{1}=2$. The reserved
SC position is assigned to $i_{5}$, and the reserved ST position
is assigned to $i_{6}$. Since two OBC positions remained vacant,
the initial seat allocation is \emph{updated} to $q_{s}^{2}=\left(5,1,1,1\right)$. 

\paragraph{\emph{Iteration }2\emph{: }}

Open-category positions are assigned to $\left\{ i_{1},i_{2},i_{3},i_{4},i_{5}\right\} $.
The only OBC candidate, $i_{4}$, now obtains open-category positions.
The previously held OBC position is now vacated. That is, $\tau^{2}=1$.
Individual $i_{6}$ is assigned to the reserved ST position. Since
$i_{7}$ and $i_{8}$ are GC individuals, the reserved SC and OBC
seats remain vacant. The vacant OBC seat is reverted to open-category.
Hence, the new distribution becomes $q_{s}^{3}=\left(6,1,1,0\right)$. 

\paragraph{\emph{Iteration 3:} }

Open-category positions are assigned to $\left\{ i_{1},i_{2},i_{3},i_{4},i_{5},i_{6}\right\} $.
Since $i_{7}$ and $i_{8}$ are GC individuals, the reserved SC and
ST seats remain unfilled. Since there is no vacant OBC position, the
procedure ends. Therefore, we have 
\[
C_{s}^{BT}\left(A,q_{s}\right)=\left\{ i_{1},i_{2},i_{3},i_{4},i_{5},i_{6}\right\} .
\]

Our first result shows that claiming affirmative action privileges
is safe for the members of reserve categories under $C^{BT}$. 
\begin{thm}
The backward transfers choice rule comply with the spirit of affirmative
action. 
\end{thm}

\subsection{DA Mechanism with Backward Transfers Choice Rules (DA-BT)}

In centralized markets with multiple institutions we propose the following
mechanism, which not only resolves the incentive issues caused by
the multi-run DA, but also provide welfare improvements. 

Let $P=\left(P_{i}\right)_{i\in\mathcal{I}}$ be the vector of\emph{
reported} preference relations and $T=\left(t_{i}\right)_{i\in\mathcal{I}}$
be the reported profile of reserve category membership of individuals.
Given the backward transfers choice function of each institution $s\in\mathcal{S}$,
$C_{s}^{BT}$---the outcome of the DA-BT mechanism is computed as
follows: 

\paragraph{Step 1. }

Each individual in $\mathcal{I}$ applies to his top choice institution.
Let $\mathcal{A}_{s}^{1}$ be the set of individuals who applies to
institution $s$, for each $s\in\mathcal{S}$. Each institution $s\in\mathcal{S}$
holds onto applicants in $C_{s}^{BT}\left(\mathcal{A}_{s}^{1},q_{s}\right)$
and rejects the rest. 

\paragraph{Step n$\protect\geq2$.}

Each individual who was rejected in the previous step applies to the
best institution that has not rejected him. Let $\mathcal{A}_{s}^{n}$
be the union of the set of individuals who were tentatively held by
institution $s$ at the end of Step $n-1$ and the set of new proposers
of $s$ in Step $n$. Each institution $s\in\mathcal{S}$ tentatively
holds individuals in $C_{s}^{BT}\left(\mathcal{A}_{s}^{n},q_{s}\right)$
and rejects the rest. 

The algorithm terminates when there is no rejection. 

There is a great benefit to re-running the choice rules rather than
the DA algorithm to revert unfilled OBC positions. Consider Example
4. The multi-run DA outcome was that individuals $i_{1}$ and $i_{3}$
receive their top choices, while individuals $i_{2}$ and $i_{4}$
are assigned to their second choices under the true preferences. The
outcome of the DA-BT for the same market is $\left(\begin{array}{cc}
a & b\\
\left\{ i_{1},i_{2}\right\}  & \left\{ i_{3},i_{4}\right\} 
\end{array}\right)$, where all individual are assigned to their top choices. By re-iterating
the choice rule rather than the DA algorithm, some unnecessary rejections
chains that occur during the multi-run DA are prevented. 

Our next result states that this observation can be generalized. 
\begin{thm}
DA-BT Pareto dominates the multi-run DA mechanism at every problem
$P$. 
\end{thm}
Moreover, the DA-BT gives individuals incentives to state their preferences
and reserve category memberships truthfully. 
\begin{thm}
DA-BT is incentive-compatible. 
\end{thm}
Theorems 2 and 3 reveal that DA-BT successfully retrieves the shortcomings
of the multi-run DA and establish a basis for a possible future reform
in admissions to technical colleges in India. 

\subsection{Characterization of the DA-BT Mechanism }

In this section, we formulate legal requirements and policy perspectives
in India as formal axioms in the $\mathbf{presence}$ of de-reservations. 

Our first axiom is a standard one. 

\paragraph{Individual rationality. }

An assignment $\eta$ is $\mathbf{individually}$ $\mathbf{rational}$
if, for every individual $i\in\mathcal{I}$, 
\[
\mu_{i}\left(\eta\right)R_{i}\emptyset.
\]
A mechanism $\varphi$ is $\mathbf{individually}$ $\mathbf{rational}$
if $\varphi\left(P,T\right)$ is individually rational for any profile
$\left(P,T\right)\in\mathcal{P}\times\mathcal{T}$. 

Individual rationality guarantees that an individual never prefers
to be unassigned to her assignment. 

Our next axiom, \emph{meritocracy}, is a natural fairness property
that says if an individual is not matched with an institution that
she prefers over her assigned institution, then all individuals who
receive either open or reserve category positions that she is eligible
for in the preferred institution must have higher priority. Baswana
et al. (2018) describe it as follows: 
\begin{quote}
\emph{``The seat allocation produced must satisfy the property that
if a candidate is denied admission to a particular program, then no
other candidate with an inferior rank in the relevant merit list should
be admitted to that program. That is, the allocation must be consistent
with a cutoff rank for each program in the relevant merit list. These
cutoff ranks are publicly announced.'' }
\end{quote}

\paragraph{Meritocracy. }

An assignment $\eta$ complies with $\mathbf{meritocracy}$ if for
every pair $\left(i,s\right)$ such that $sP_{i}\mu(i)$, for all
$j$ such that $\left(j,o\right)\in\eta\left(s\right)$ or $\left(j,t(i)\right)\in\eta(s)$
where $t(i)\in\mathcal{R}$,  we have $j\succ_{s}i$. 

A mechanism $\varphi$ complies with $\mathbf{meritocracy}$ if $\varphi\left(P,T\right)$
complies with meritocracy for any profile $\left(P,T\right)\in\mathcal{P}\times\mathcal{T}$. 

Our next axiom is also a standard one. It is a mild efficiency requirement. 

\paragraph{Non-wastefulness. }

An assignment $\eta$ is $\mathbf{non-wasteful}$ if for every pair
$\left(i,s\right)$ such that $sP_{i}\mu(i)$, all positions $i$
is eligible at $s$ must be exhausted. That is, 
\[
\mid\left\{ j\mid\left(j,t(i)\right)\in\eta(s)\right\} \mid=q_{s}^{t(i)},
\]
for all $t(i)\in\mathcal{R}$, and also 
\[
\mid\left\{ j\mid\left(j,o\right)\in\eta(s)\;or\;(j,OBC)\in\eta(s)\right\} \mid=q_{s}^{o}+q_{s}^{OBC}.
\]

A mechanism $\varphi$ is $\mathbf{non-wasteful}$ if $\varphi\left(P,T\right)$
is non-wasteful for any profile $\left(P,T\right)\in\mathcal{P}\times\mathcal{T}$. 

It simply says that if an individual prefers an institution to her
assigned institution, then all categories in the preferred institution
she is eligible for must be exhausted. Note that the second condition
takes de-reservations into account as follows: at the preferred institution,
not only $q_{s}^{o}$ open-category positions but also $q_{s}^{OBC}$
positions must be allocated either as open-category or OBC category
because any unfilled OBC position must be provided as an open-category
position. 

Our next axiom formulates the \emph{over-and-above principle} \emph{in
the presence of de-reservation}. In the absence of de-reservations,
the over-and-above approach says that open-category positions taken
by reserve category applicants from the initially set $q_{s}^{o}$
positions are not counted against their respective reservations. It
is executed by filling open-category positions before reserve category
positions. The following quote from the ''\emph{Business Rules for
Joint Seat Allocation for the Academic Programs offered by the IITs,
NITs, IIESTs, IIITs, and Other-GFTIs for the academic year 2021-22}''
clearly describe the order at which positions are filled in the presence
of de-reservations: 
\begin{quotation}
\emph{``The justification for the above order is the following: The
candidate is eligible for OPEN seats. So attempts must be made to
allocate her a OPEN seat before her own category}.''
\end{quotation}
Both $q_{s}^{o}$ (initially set) open-category positions and \emph{reverted
surplus OBC positions }should be allocated before allocating the reserve
category positions. This observation motivates us to extend the over-and-above
principle---which is defined in the absence of de-reservations---so
that reserve category applicants who took positions from the reverted
surplus OBC positions are also not counted against their respective
reservations. 

Our next axiom also reflects the policymakers' goal of having higher
cutoff scores for open-category poistions compared to reserved category
positions. Note that modifying OBC positions' priorities would not
achieve this goal. 
\begin{description}
\item [{Open-first.}] An assignment $\eta$ satisfies $\mathbf{open-first}$
if, 
\item [{(1)}] $\mid\left\{ j\mid\left(j,o\right)\in\eta(s)\;or\;\left(j,OBC\right)\in\eta(s)\right\} \mid=min\left\{ \mid\eta(s)\mid,q_{s}^{o}+q_{s}^{OBC}\right\} $
holds for every $s$, 
\item [{(2)}] for all $i,j\in\mathcal{I}$ such that $\left(i,o\right)\in\eta(s)$
and $(j,r)\in\eta(s)$ where $r\in\mathcal{R}$, we have $i\succ_{s}j,$
and 
\item [{(3)}] for any $i$ such that $\left(i,o\right)\in\eta(s)$ and
$t(i)\in\mathcal{R}$, $\eta^{'}(s)=\left(\eta(s)\setminus\left\{ \left(i,o\right)\right\} \right)\bigcup\left\{ i,t(i)\right\} $
violates either $(1)$ or $(2)$.
\end{description}
The open-first property aims to allocate open-category positions to
high-scoring applicants while taking de-reservation policy into account
and preventing unnecessary de-reservations. Condition (1) ensures
that unfilled OBC positions are made open-category positions. Condition
(2) says that every individual with an open-category position must
have a higher priority than every individual with a reserved category
position. This is true not only for initially set open-category positions
but also for the ones that are de-reserved from the OBC. Finally,
condition (3) prevents unnecessary de-reservations from OBC to open-category.
Note that, under every assignment that satisfies the open-first axiom,
open-category cutoff score is higher than the cutoff scores of reserve
categories. 

A mechanism $\varphi$ satisfies $\mathbf{open-first}$ if $\varphi\left(P,T\right)$
satisfies open-first for any profile $\left(P,T\right)\in\mathcal{P}\times\mathcal{T}$. 

The multi-run DA mechanism is an attempt to satisfy the open-first
principle in the presence of de-reservation; however, it has unintended
consequences. The DA-BT mechanism satisfies the open-first property
and corrects the flaws of the multi-run DA. 

We are now ready to present our characterization result. 
\begin{thm}
Fix a profile of priority orders $\left(\succ_{s}\right)_{s\in\mathcal{S}}$.
A mechanism $\varphi$ satisfies 

(1) individual rationality, 

(2) meritocracy, 

(3) non-wastefulness, 

(4) open-first, and 

(5) incentive-compatibility, 

if and only if $\varphi$ is the DA-BT mechanism. 
\end{thm}

\section{Related Literature}

There are four strands of literature directly related to the current
paper.

\paragraph{1. Affirmative action in India.}

Echenique and Yenmez (2015) is the first paper that discusses affirmative
action in India from a market design perspective and provides college
admission in India as an example of controlled school choice in Appendix
C.1. Aygün and Turhan (2017) discuss issues in admissions to IITs.
Both papers consider \emph{vertical reservations only}. Aygün and
Turhan (2020) formulate vertical reservations and de-reservations
in admissions to technical colleges in India. 

There are also \emph{horizontal reservations} implemented within each
reserved category, including the open category. This manuscript does
not model horizontal reservations for simplicity and focuses only
on the joint implementation of vertical reservations and OBC de-reservations.
Sönmez and Yenmez (2021) formulate vertical and horizontal reservations
jointly and relates Indian laws on reservation policy to matching
theory. However, Sönmez and Yenmez (2021) do not model OBC de-reservations
and, hence, restricts attention to allocating government jobs, where
OBC de-reservation is not mandated. While this restriction simplifies
their analysis considerably, the de-reservation policy is critical
for admissions to technical colleges due to the landmark SCI ruling
in 2008. The current paper also deviate significantly from Sönmez
and Yenmez (2021) in that this paper considers both priority design
for a single institution and mechanism design for centralized marketplaces
with multiple institutions, while Sönmez and Yenmez (2021) only consider
priority design for a single institution. 

Building on Aygün and Turhan (2020), Aygün and Turhan (2022) formulate
vertical reservations, horizontal reservations, and de-reservations
all together. Aygün and Turhan (2022) present significant evidence
that many reserve category members consider open-category and reserve-category
positions differently and model individuals' preferences over institution-position
category pairs for reserve category members. Similar to Aygün and
Turhan (2020), Aygün and Turhan (2022) invoke the matching with contracts
framework to model allocation problems in India. In this paper, we
model individuals' preferences only over institutions. There are two
reasons why we chose this modeling approach. First, we want to focus
on the joint implementation issue of reservations and de-reservations
only and keep the other aspects of the problem as simple as possible.
The second one is that currently, individuals are only asked to rank
institutions in allocating public jobs and public school seats in
India, even though reporting category membership is optional. The
framework of this paper can be straightforwardly extended to this
larger preference domain and modeled via matching with contracts. 

This paper discusses the joint implementation of reservation and de-reservations
via the multi-run DA mechanism of Baswana et al. (2018 and 2019) and
reveal its drawbacks. We offer an alternative mechanism that fixes
these failures. 

Thakur (2020) studies a job allocation problem in Indian Administrative
Services, where the de-reservation policy is not implemented. Thakur
(2020) does not model horizontal reservations. 

Similar to ours, Sönmez and Yenmez (2022) characterize their suggested
mechanism via axioms reflecting policy goals in India, but the authors
assume away de-reservations altogether. 

\paragraph{2. Characterizations of the DA mechanisms. }

There is important literature on the characterization of the deferred
acceptance and cumulative offer mechanisms. Balinski and Sönmez (1999)
is the first paper to characterize deferred acceptance in the student
placement context. Hirata and Kasuya (2017) and Hatfield, Kominers,
and Westkamp (2021) characterize the cumulative offer mechanism. Both
papers take exogenously given choice rules for institutions that satisfy
certain mathematical conditions. Recently, Greenberg, Pathak, and
Sönmez (2021) provided a characterization for the cumulative offer
mechanism under an endogenous choice rule for branches that echos
the policy goals of the US Army in a cadet-branch matching framework.
Our characterization of the deferred acceptance is similar to theirs
in that our choice rule is endogenous and reflects the policy perspective
of Indian authorities. 

\paragraph{3. Reserve policies. }

There is a large literature on reserve policies in different contexts.
Examples include (1) Hafal\i r, Yenmez, and Yildirim (2013), Westkamp
(2013), Echenique and Yenmez (2015), Do\u{g}an (2016 and 2017), Dur
et al. (2018), Dur, Pathak and Sönmez (2020), Correa et al. (2019),
Aygün and Bó (2021), Do\u{g}an and Y\i ld\i z (2021), and Abdulkadiro\u{g}lu
and Grigoryan (2021) in school choice context; Pathak et al. (2020),
Aziz and Brandl (2021), and Grigoryan (2021) in allocation of medical
resources; Pathak, Rees-Jones and Sönmez (2020) in allocation of H-1B
visa allocation in the US, Gonczarowski et al. (2020) in Machinot
gap year program in Israel. Our theoretical analysis deviates from
this literature by concurrently implementing reservation and de-reservation
policies. 

This paper introduces the BT choice rule for institutions. Aygün and
Turhan (2022b) provide the simpler choice rule that is outcome equivalent
to the BT choice rule. The simpler choice rule interprets the OBC
reservation as a \emph{soft reserve} and modifies the priorities of
OBC positions to make unfilled OBC positions available to others.
However, the BT choice rule has a crucial advantage compared to this
simpler choice rule via soft reserves in that the BT choice rule guarantees
that the open-category cutoff score is higher than the OBC cutoff
after unfilled OBC positions are provided to others. 

\paragraph{4. Matching with diversity considerations. }

There is also a large literature on diversity constraints in matching
problems. Important work in this line of literature include Abdulkadiro\u{g}lu
and Sönmez (2003), Abdulkadiro\u{g}lu (2005), Kojima (2012), Ehlers
et al. (2014), Kamada and Kojima (2015 and 2017),\emph{ }Kominers
and Sönmez (2016), Kurata et al. (2017), Fragiadakis and Troyan (2017),
Kojima, Tamura, and Yokoo (2018), Nguyen and Vohra (2019), Avataneo
and Turhan (2020), Imamura (2020), Aygün and Bó (2021), Delacrétaz
(2021), and Aziz, Baychkov, and Biró (2021), among others. 

\section{Conclusion}

This paper studies the joint implementation of reservation and de-reservation
policies in the context of Indian technical college admissions. We
discuss the unintended consequences of the de-reservation policy implemented
since 2015. We introduce a new idea to implement de-reservations via
institutions' choice functions. To this end, we design a novel choice
procedure by taking legal requirements and policy perspectives of
Indian policymakers. We propose the DA mechanism with respect to these
choice rules and show that it is incentive-compatible for applicants,
and Pareto improves upon the currently implemented mechanism. More
importantly, we define legal requirements and policy goals as formal
axioms and show that the DA-BT mechanism is the only mechanism achieving
these objectives.

\section{APPENDIX (FOR ONLINE PUBLICATION)}

\paragraph{Proof of Theorem 1. }

Suppose that individual $i$---who is a member of category $r\in\mathcal{R}$---is
not chosen by $C_{s}^{BT}\left(\cdot,q_{s}\right)$ when she reports
her membership to category $r$, that is $t_{i}=r$. We need to show
that she is not chosen if she reports $t_{i}^{'}=GC$. $i$ being
not chosen when she reports $t_{i}=r$ means $i$ gets rejected for
open-category positions in every iteration of $C_{s}^{BT}$. If $i$
reports $t_{i}^{'}=GC$, then she gets rejected for open-category
positions in every iteration. This is because she cannot change the
set of applicants who apply for open-category positions and the number
of unfilled OBC seats that are reverted to open-category at the end
of each iteration. 

\paragraph{Proof of Theorem 3. }

We first prove the following lemma that will be useful in the proof
of our theorem. We introduce the necessary notation first. Consider
a set of applicants $A\subseteq\mathcal{I}$. Let $A^{SC}$, $A^{ST}$,
and $A^{OBC}$ be sets of individuals who belong to SC, ST, and OBC,
respectively, in the set $A$. In the backward transfers choice rules,
let $q_{s}=q_{s}^{1}$ be the initial vector of capacities of categories,
$N$ be the last iteration, and $q_{s}^{N}$ denote the vector of
capacities of categories at institution $s\in\mathcal{S}$ in the
last iteration of $C_{s}^{BT}$. By definition, $C_{s}^{BT}\left(A,q_{s}\right)=C^{IN}\left(A,q_{s}^{N}\right)$.
We denote by $C_{s}^{o}\left(A,\left(q_{s}^{n}\right)^{o}\right)$
the set of individuals chosen from the open-category given a set of
applicants $A$ , and capacity $\left(q_{s}^{n}\right)^{o}$ of the
open-category at iteration $n$ of $C_{s}^{BT}$ . This choice rule
selects applicants following the priority ordering $\succ_{s}$ of
institution $s$ up to the capacity $\left(q_{s}^{n}\right)^{o}$. 
\begin{lem}
Given a set of applicants $A\subseteq\mathcal{I}$ and a vector of
initial distribution of positions over categories $q_{s}$ of institution
$s\in\mathcal{S}$, $N$ is the last iteration of the backward transfers
choice rule $C_{s}^{BT}\left(A,q_{s}\right)$ if, and only if, either
one of the following holds: 

$(1)$ $\mid\left(C_{s}^{o}(A,(q_{s}^{N})^{o})\setminus C_{s}^{o}(A,(q_{s}^{1})^{o})\right)\setminus A^{OBC}\mid=\tau^{1}$,
where $\tau^{1}$ is the number of unfilled OBC positions at the end
of iteration 1, and $A^{OBC}=\left\{ i\in A\mid t_{i}=OBC\right\} $. 

$(2)$ $\left(q_{s}^{N}\right)^{o}=\left(q_{s}^{1}\right)^{o}+\left(q_{s}^{1}\right)^{OBC}$. 
\end{lem}

\paragraph{Proof of Lemma 1. }

We will first show that given a set of applicants $A\subseteq\mathcal{I}$
and a vector of initial distribution of positions over categories
$q_{s}$ of institution $s\in\mathcal{S}$, $N$ is the $\mathbf{last}$
iteration of $C_{s}^{BT}\left(A,q_{s}\right)$ if, and only if, either
one of the following holds: 
\begin{enumerate}
\item $\mid\left(C_{s}^{o}(A,(q_{s}^{N})^{o})\setminus C_{s}^{o}(A,(q_{s}^{1})^{o})\right)\setminus A^{OBC}\mid=\tau^{1}$,
where $\tau^{1}$ is the number of unfilled OBC positions at the end
of iteration 1, and $A^{OBC}=\left\{ i\in A\mid t_{i}=OBC\right\} $. 
\item $\left(q_{s}^{N}\right)^{o}=\left(q_{s}^{1}\right)^{o}+\left(q_{s}^{1}\right)^{OBC}$. 
\end{enumerate}
$\left(\Leftarrow\right)$ If we have $\left(q_{s}^{N}\right)^{o}=\left(q_{s}^{1}\right)^{o}+\left(q_{s}^{1}\right)^{OBC}$,
then it means the number of vacant OBC seats at iteration $N$ is
0. Hence, $N$ is the final iteration. Now suppose $\mid\left(C_{s}^{o}(A,(q_{s}^{N})^{o})\setminus C_{s}^{o}(A,(q_{s}^{1})^{o})\right)\setminus A^{OBC}\mid=\tau^{1}$.
First, note that the following equality holds in iteration 1:
\[
\left(q_{s}^{1}\right)^{o}+\left(q_{s}^{1}\right)^{OBC}=\tau^{1}+\mid A^{OBC}\cup C_{s}^{o}\left(A,\left(q_{s}^{1}\right)^{o}\right)\mid.
\]

Since the total number of OBC and open-category positions in every
iteration remains unchanged, we have the following equality holding
at iteration $N$: 
\[
\left(q_{s}^{1}\right)^{o}+\left(q_{s}^{1}\right)^{OBC}=\left(q_{s}^{N}\right)^{o}+\left(q_{s}^{N}\right)^{OBC}=\tau^{N}+\mid A^{OBC}\cup C_{s}^{o}\left(A,\left(q_{s}^{N}\right)^{o}\right)\mid.
\]

Since $\mid\left(C_{s}^{o}(A,(q_{s}^{N})^{o})\setminus C_{s}^{o}(A,(q_{s}^{1})^{o})\right)\setminus A^{OBC}\mid=\tau^{1}$,
we have 
\[
\mid C_{s}^{o}\left(A,\left(q_{s}^{N}\right)^{o}\right)\setminus\left(C_{s}^{o}\left(Y,\left(q_{s}^{1}\right)^{o}\right)\cup A^{OBC}\right)\mid=\tau^{1},
\]
which implies 
\[
\mid A^{OBC}\cup C_{s}^{o}\left(A,\left(q_{s}^{N}\right)^{o}\right)\mid-\mid A^{OBC}\cup C_{s}^{o}\left(A,\left(q_{s}^{1}\right)^{o}\right)\mid=\tau^{1}.
\]
 Therefore, we have 
\[
\mid A^{OBC}\cup C_{s}^{o}\left(A,\left(q_{s}^{N}\right)^{o}\right)\mid=\tau^{1}+\left(q_{s}^{1}\right)^{o}+\left(q_{s}^{1}\right)^{OBC}-\tau^{1}=\left(q_{s}^{N}\right)^{o}+\left(q_{s}^{N}\right)^{OBC},
\]
 which implies $\tau^{N}=0$. Thus, $N$ is the last iteration.

$\left(\Rightarrow\right)$ Let $N$ be the last iteration of $C_{s}^{BT}\left(A,q_{s}\right)$.
Toward a contradiction, suppose that neither $(1)$ nor $(2)$ holds.
That is, in the final step 
\[
\left(q_{s}^{N}\right)^{o}<\left(q_{s}^{1}\right)^{o}+\left(q_{s}^{1}\right)^{OBC}\Longrightarrow\left(q_{s}^{N}\right)^{OBC}>0,
\]
 and 
\[
\mid\left(C_{s}^{o}(A,(q_{s}^{N})^{o})\setminus C_{s}^{o}(A,(q_{s}^{1})^{o})\right)\setminus A^{OBC}\mid\neq\tau^{1},
\]
 which implies 
\[
\mid A^{OBC}\cup C_{s}^{o}\left(A,\left(q_{s}^{N}\right)^{o}\right)\mid-\mid A^{OBC}\cup C_{s}^{o}\left(A,\left(q_{s}^{1}\right)^{o}\right)\mid\neq\tau^{1}.
\]
 Thus, we have 
\[
\mid A^{OBC}\cup C_{s}^{o}\left(A,\left(q_{s}^{N}\right)^{o}\right)\mid\neq\tau^{1}+\left(q_{s}^{1}\right)^{o}+\left(q_{s}^{1}\right)^{OBC}-\tau^{1}=\left(q_{s}^{1}\right)^{o}+\left(q_{s}^{1}\right)^{OBC}=\left(q_{s}^{N}\right)^{o}+\left(q_{s}^{N}\right)^{OBC}.
\]

This implies that 
\[
\mid A^{OBC}\setminus C_{s}^{o}\left(A,\left(q_{s}^{N}\right)^{o}\right)\mid+\mid C_{s}^{o}\left(Y,\left(q_{s}^{N}\right)^{o}\right)\mid\neq\left(q_{s}^{N}\right)^{o}+\left(q_{s}^{N}\right)^{OBC}.
\]
 Since $C_{s}^{o}$ is a q-responsive choice function, we have two
cases to consider:

\paragraph{Case 1: $\mid C_{s}^{o}\left(A,\left(q_{s}^{N}\right)^{o}\right)\mid<\left(q_{s}^{N}\right)^{o}.$}

In this case, all individuals are accepted by $C_{s}^{o}\left(A,\left(q_{s}^{N}\right)^{o}\right)$.
Hence, 
\[
A^{OBC}\setminus C_{s}^{o}\left(A,\left(q_{s}^{N}\right)^{o}\right)=\emptyset.
\]
This implies $\tau^{N}=\left(q_{s}^{N}\right)^{OBC}>0$. That means
$N$ is not the final iteration. This is a contradiction. 

\paragraph{Case 2: $\mid C_{s}^{o}\left(A,\left(q_{s}^{N}\right)^{o}\right)\mid=\left(q_{s}^{N}\right)^{o}.$}

In this case, 
\[
\mid A^{OBC}\setminus C_{s}^{o}\left(A,\left(q_{s}^{N}\right)^{o}\right)\mid\neq\left(q_{s}^{N}\right)^{OBC},
\]
since $N-1$ is not the final iteration of $C_{s}^{BT}\left(A,q_{s}\right)$
by construction, i.e., $\tau^{N-1}>0$, we have 

\[
\begin{array}{c}
(i)\;C_{s}^{o}\left(A,\left(q_{s}^{N-1}\right)^{o}\right)\subset C_{s}^{o}\left(A,\left(q_{s}^{N}\right)^{o}\right)\\
(ii)\;\mid A^{OBC}\setminus C_{s}^{o}\left(A,\left(q_{s}^{N-1}\right)^{o}\right)\mid=\left(q_{s}^{N-1}\right)^{OBC}-\tau^{N-1}=\left(q_{s}^{N}\right)^{OBC}
\end{array}
\]
 $(i)$ and $(ii)$ imply 

\[
A^{OBC}\setminus C_{s}^{o}\left(A,\left(q_{s}^{N}\right)^{o}\right)\subset A^{OBC}\setminus C_{s}^{o}\left(A,\left(q_{s}^{N}\right)^{o}\right)
\]
 and 
\[
\mid A^{OBC}\setminus C_{s}^{o}\left(A,\left(q_{s}^{N}\right)^{o}\right)\mid\leq\mid A^{OBC}\setminus C_{s}^{o}\left(A,\left(q_{s}^{N-1}\right)^{o}\right)\mid=\left(q_{s}^{N}\right)^{OBC}.
\]

Then, by $\mid A^{OBC}\setminus C_{s}^{o}\left(A,\left(q_{s}^{n}\right)^{o}\right)\mid\neq\left(q_{s}^{N}\right)^{OBC}$,
we have 
\[
\mid A^{OBC}\setminus C_{s}^{o}\left(A,\left(q_{s}^{N}\right)^{o}\right)\mid<\left(q_{s}^{N}\right)^{OBC},
\]
 which implies that $\tau^{N}>0$. Hence, $N$ is not the final iteration.
This is a contradiction and ends the proof of Lemma 1. 

We need to show that for every profile $(P,T)\in\mathcal{P}\times\mathcal{T}$,
and for each individual $i\in\mathcal{I}$, and for any possible deviation
$\left(\widetilde{P}_{i},\widetilde{t}_{i}\right)$, we have $\varphi\left(P,T\right)R_{i}\varphi\left(\left(\widetilde{P}_{i},\widetilde{t}_{i}\right),\left(P_{-i},T_{-i}\right)\right)$. 

\paragraph{(i) }

We first show that for a given profile of category membership $T=\left(t_{i}\right)_{i\in I}$,
DA-BT cannot be manipulated via preference misreporting. We prove
this by showing that $C^{BT}$ satisfies substitutability and size
monotonicity.

\paragraph{Substitutability. }

Consider $i,j\in\mathcal{I}$ and $A\subset\mathcal{I}\setminus\{i,j\}$
such that $i\notin C_{s}^{BT}\left(A\cup\left\{ i\right\} \right)$.
We need to show that $i\notin C_{s}^{BT}\left(A\cup\left\{ i,j\right\} \right)$. 

Let $\tau^{k}$ and $\widetilde{\tau}^{k}$ denote the number of vacant
OBC positions at the end of iteration $k$ under $C_{s}^{BT}\left(A\cup\{i\},q_{s}\right)$
and $C_{s}^{BT}\left(A\cup\{i,j\},q_{s}\right)$, respectively. Let
$N$ and $\widetilde{N}$ be the last steps of $C_{s}^{BT}\left(A\cup\{i\},q_{s}\right)$
and $C_{s}^{BT}\left(A\cup\{i,j\},q_{s}\right)$, respectively. Note
that, by Lemma 1, $\widetilde{N}\leq N$ and $\left(q_{s}^{N}\right)^{o}\geq\left(q_{s}^{\widetilde{N}}\right)^{o}$,
where $\left(q_{s}^{N}\right)^{o}$ and $\left(q_{s}^{\widetilde{N}}\right)^{o}$
are the capacities of open-category at the last steps of $C_{s}^{BT}\left(A\cup\{i\},q_{s}\right)$
and $C_{s}^{BT}\left(A\cup\{i,j\},q_{s}\right)$, respectively. Let
$A_{r}\subseteq A\cup\{i\}$ denotes the set of individuals who belong
to reserve category $r\in\mathcal{R}$. For each $r\in\mathcal{R}$,
define $A_{r}^{'}=A_{r}\setminus C_{s}^{o}\left(A\cup\{i\},\left(q_{s}^{N}\right)^{o}\right)$. 

If $i\notin C_{s}^{BT}\left(A\cup\left\{ i\right\} ,q_{s}\right)$,
then we know that $i$ is not in the top $\left(q_{s}^{N}\right)^{o}$
in the set $A\cup\{i\}$. This implies that $i$ is not in top $\left(q_{s}^{\widetilde{N}}\right)^{o}$
in the set $A\cup\{i,j\}$. So, $i$ cannot be chosen for an open-category
position from $A\cup\{i,j\}$. 

We now show that $i$ cannot be chosen for a reserve category $t_{i}\in\mathcal{R}$
position. First, suppose that $t_{i}=OBC$. Since $i$ was not chosen
for an OBC position from the set $A\cup\{i\}$, and when $j$ is added
to the set $A\cup\{i\}$, $i$ cannot be chosen for an OBC position
because adding $j$ (weakly) increases the competition for OBC positions. 

Now, suppose that $t_{i}\in\{SC,ST\}$. The capacities of reserved
SC and ST categories are the same for the choice processes starting
with $A\cup\{i\}$ and $A\cup\{i,j\}$ in every iteration of the $C_{s}^{BT}$.
Moreover, we have 
\[
A\cup\{i,j\}\setminus C_{s}^{o}\left(A\cup\{i,j\},\left(q_{s}^{\widetilde{N}}\right)^{\mathbf{t}(i)}\right)\supseteq A\cup\{i\}\setminus C_{s}^{o}\left(A\cup\{i\},\left(q_{s}^{N}\right)^{o}\right),
\]
 for both $t_{i}=SC$ and $t_{i}=ST$. That is, the competition for
the SC and ST positions will be (weakly) higher in the choice process
starting with $A\cup\{i,j\}$ than the choice process starting with
$A\cup\{i\}$. Since $i$ was not chosen for reserved $t_{i}$ position
from $A\cup\{i\}\setminus C_{s}^{o}\left(A\cup\{i\},\left(q_{s}^{N}\right)^{o}\right)$,
we can conclude that $i$ will not be chosen for reserved $t_{i}$
position from $A\cup\{i,j\}\setminus C_{s}^{o}\left(A\cup\{i,j\},\left(q_{s}^{\widetilde{N}}\right)^{\mathbf{t}(i)}\right)$.
Therefore, $i$ cannot be chosen for reserved $t_{i}\in\mathcal{R}$
positions. This ends our proof of substitutability. 

\paragraph{Size monotonicity. }

Consider $i\in\mathcal{I}$ and $A\subseteq\mathcal{I}\setminus\left\{ i\right\} $.
We need to show that $\mid C_{s}^{BT}\left(A\right)\mid\leq\mid C_{s}^{BT}\left(A\cup\left\{ i\right\} \right)\mid$.
We consider following two cases: 

\subparagraph{Case 1. $\mid A\mid\protect\leq q_{s}^{o}+q_{s}^{OBC}$.}

In this case, all individuals in $A$ will be chosen. When $i$ is
added to the set $A$, the number of chosen individuals increases
by one and becomes $\mid A\mid+1$, if $\mid A\mid<q_{s}^{o}+q_{s}^{OBC}$.
When $\mid A\mid\leq q_{s}^{o}+q_{s}^{OBC}$, since the number of
chosen individuals is $\mid A\mid$, adding $i$ to the set $A$ does
not change the number of chosen individuals. 

\subparagraph{Case 2. $\mid A\mid>q_{s}^{o}+q_{s}^{OBC}$.}

The backward transfers choice rule $C_{s}^{BT}$ selects $min\left\{ \mid A\mid,\overline{q}_{s}\right\} $
individuals, unless either $\mid A_{SC}^{'}\mid<q_{s}^{SC}$ or $\mid A_{ST}^{'}\mid<q_{s}^{ST}$,
where $A_{r}^{'}=A_{r}\setminus C_{s}^{o}\left(A,\left(q_{s}^{N}\right)^{o}\right)$
for $r\in\left\{ SC,ST\right\} $. Note that $N$ represents the last
iteration of the backward transfers choice rule. In other words, the
choice rule $C_{s}^{BT}$ behaves as a q-responsive choice function
if the number of remaining SC and ST individuals after open-category
positions are filled are at least as many as the number of reserved
SC and ST positions, respectively. Therefore, when both $\mid A_{SC}^{'}\mid\geq q_{s}^{SC}$
or $\mid A_{ST}^{'}\mid\geq q_{s}^{ST}$, the number of chosen individuals
will be $\overline{q}_{s}$ and adding individual $i$ to the set
$A$ does not change the number of chosen individuals. When either
$\mid A_{SC}^{'}\mid<q_{s}^{SC}$ or $\mid A_{ST}^{'}\mid<q_{s}^{ST}$,
the number of chosen individuals from $A\cup\{i\}$ either stays the
same or increases by one. 

Substitutability and size monotonicity of $C^{BT}$ imply strategy-proofness
of the DA-BT by Hatfield and Milgrom (2005) given a profile of category
memberships of individuals. 

\paragraph{(ii) }

Next, we prove that, for a given preference profile $P=\left(P_{i}\right)_{i\in I}$,
DA-BT cannot be manipulated by not reporting reserve category membership. 

We adapt the following definition from Definition 8 of Afacan (2017):
A choice rule $C_{s}^{'}$ is an \emph{improvement} over a choice
rule $C_{s}$ for individual $i$ if, for any set of individuals $A$
(i) if $i\in C_{s}\left(A\right)$, then $i\in C_{s}^{'}\left(A\right)$,
and (ii) if $i\notin C_{s}\left(A\right)\cup C_{s}^{'}\left(A\right)$,
then $C_{s}\left(A\right)=C_{s}^{'}\left(A\right)$. 

Consider a reserve category $r\in\mathcal{R}$ member $i$. Let $C_{s}^{BT}$
and $\widetilde{C}_{s}^{BT}$ be backward transfers choice rules individual
$i$ faces when she does \emph{not} report and reports, respectively,
her category $r$ membership. Theorem 2 states that $\widetilde{C}_{s}^{BT}$
is an improvement over $C_{s}^{BT}$ for $i$ according to the given
definition of improvement. We say that mechanism $\psi$ \emph{respects
improvements} if for any problem $(P,C)$ and $C^{'}$ such that $C^{'}$
is an improvement over $C$ for individual $i$, $\psi(P,C^{'})R_{i}\psi(P,C)$.
Theorem 2 of Afacan (2017)\footnote{Theorem 2 of Afacan (2017) states that the generalized DA respects
improvements if choice rules of institutions satisfy the \emph{unilateral
substitutes} of Hatfield and Kojima (2010), the \emph{irrelevance
of rejected contracts}, and the \emph{size monotonicity}. $C^{BT}$
satisfies substitutability, and hence, unilateral substitutability.
Moreover, $C^{BT}$ satisfy size monotonicity, which---in conjunction
with substitutability---implies the irrelevance of rejected contracts
condition.} holds in our setting with $C^{BT}$. Then, when individuals report
their reserved category membership, they can never be hurt under DA-BT. 

From part (i), we have 
\[
\mu_{i}[\varphi\left(P,T\right)]R_{i}\mu_{i}[\varphi\left((\widetilde{P}_{i},P_{-i}),T\right)].
\]
 From part (ii), we obtain 
\[
\mu_{i}[\varphi\left((\widetilde{P}_{i},P_{-i}),T\right)]R_{i}\mu_{i}[\varphi\left((\widetilde{P}_{i},P_{-i}),(\widetilde{t}_{i},T_{-i})\right)].
\]
Transitivity of the relation $R_{i}$ gives the desired conclusion
\[
\mu_{i}[\varphi\left(P,T\right)]R_{i}\mu_{i}[\varphi\left((\widetilde{P}_{i},P_{-i}),(\widetilde{t}_{i},T_{-i})\right)].
\]

\paragraph{Proof of Theorem 2.}

We prove Theorem 2 by showing that the outcome of the multi-run DA
at a given preference profile $P=\left(P_{i}\right)_{i\in\mathcal{I}}$
is stable with respect to the backward transfers choice rules of institutions
$\left(C_{s}^{BT}\right)_{s\in\mathcal{S}}$ at the same preference
profile $P$ given the profile of individuals' reserve category membership
profile. We have shown that $C^{BT}$ is substitutable and size monotonic
in the proof of Theorem 3 above.  Therefore, by Theorem 4 of Hatfield
and Milgrom (2005), the DA outcome is the \emph{individual-optimal
}stable outcome, where stability is defined with respect to a profile
of backward transfers choice rules $\left(C_{s}^{BT}\right)_{s\in\mathcal{S}}$.
That is, each applicant weakly prefers the outcome of the generalized
DA to her assignment in any other stable matching. 

Let $v$ be the outcome of the multi-run DA at preference profile
$P=\left(P_{i}\right)_{i\in\mathcal{I}}$. That is, $v=\Phi\left(P,q^{L}\right)$,
where $L$ denotes the last iteration of the DA in multi-run DA algorithm.
We will show that $v$ is stable with respect to the profile of backward
transfers choice rules $\left(C_{s}^{BT}\right)_{s\in\mathcal{S}}$.
A matching $\mu$ is $\mathbf{stable}$ with respect to the profile
of applicants' preferences $P=\left(P_{i}\right)_{i\in\mathcal{I}}$
and a profile backward transfers choice rules of institutions $C^{BT}=\left(C_{s}^{BT}\right)_{s\in\mathcal{S}}$
if, 
\begin{enumerate}
\item for every individual $i\in\mathcal{I}$, $\mu(i)R_{i}\emptyset$, 
\item for every institution $s\in\mathcal{S}$, $C_{s}(\mu(s))=\mu(s)$,
and 
\item there is no $(i,s)$ such that $sP_{i}\mu(i)$ and $i\in C_{s}(\mu(s)\cup\{i\})$.
\end{enumerate}

\paragraph{Individual Rationality for Individuals. }

Since the preference profile in multi-run DA and DA-BT are the same,
for every individual $i\in\mathcal{I}$, $v(i)R_{i}\emptyset$. 

\paragraph{Individual Rationality for Institutions. }

We need to show that the outcome of the multi-run DA at preference
profile $P$ is individually rational for institution $s\in\mathcal{S}$
with respect to its backward transfers choice rule $C_{s}^{BT}$,
for all institutions $s\in\mathcal{S}$. That is, $C_{s}^{BT}(v\left(s\right))=v\left(s\right)$
for all institutions $s\in\mathcal{S}$, where $v(s)$ denotes the
set of applicants who are matched to institution $s$ under the multi-run
DA. 

We will first prove a lemma that will be the key to prove individual
rationality for institutions. Consider a set of applicants $A\subseteq\mathcal{I}$.
Let $A^{SC}$, $A^{ST}$, and $A^{OBC}$ be sets of individuals who
belong to SC, ST, and OBC, respectively, in the set $A$. Let $q_{s}=q_{s}^{1}$
be the initial vector of capacities of categories, $N$ be the last
iteration , and $q_{s}^{N}$ denote the vector of capacities of categories
at institution $s\in\mathcal{S}$ in the last iteration of $C_{s}^{BT}$.
By definition, $C_{s}^{BT}\left(A,q_{s}\right)=C^{IN}\left(A,q_{s}^{N}\right)$.
Let $C_{s}^{o}\left(A,\left(q_{s}^{n}\right)^{o}\right)$ denote the
set of individuals chosen from the open-category given a set of applicants
$A$ and the capacity $\left(q_{s}^{n}\right)^{o}$ of the open-category
at iteration $n$ of $C_{s}^{BT}$. This choice rule selects applicants
following the priority ordering $\succ_{s}$ of institution $s$ up
to the capacity $\left(q_{s}^{n}\right)^{o}$. 

By Lemma 1, $C_{s}^{BT}\left(A,q_{s}\right)$ is finalized as soon
as either one of the conditions in Lemma 1 is satisfied. In the last
iteration, call it $L$, of the multi-run DA we have either 

\[
\mid C_{s}^{o}\left(A,\left(q_{s}^{L}\right)^{o}\right)\cup A^{OBC}\mid=\left(q_{s}^{1}\right)^{o}+\left(q_{s}^{1}\right)^{OBC}
\]
 or 
\[
\left(q_{s}^{L}\right)^{o}=\left(q_{s}^{1}\right)^{o}+\left(q_{s}^{1}\right)^{OBC}.
\]
 Then, we get 
\[
\left(q_{s}^{L}\right)^{o}\geq\left(q_{s}^{n}\right)^{0},
\]
which implies 
\[
C_{s}^{o}\left(A,\left(q_{s}^{n}\right)^{o}\right)\subseteq C_{s}^{o}\left(A,\left(q_{s}^{L}\right)^{o}\right),
\]
 which, in turn, implies 
\[
A^{OBC}\cup C_{s}^{o}\left(A,\left(q_{s}^{n}\right)^{o}\right)\subseteq A^{OBC}\cup C_{s}^{o}\left(A,\left(q_{s}^{L}\right)^{o}\right).
\]
 Hence, we obtain 
\[
\mid A^{OBC}\cup C_{s}^{o}\left(A,\left(q_{s}^{n}\right)^{o}\right)\mid\leq\mid A^{OBC}\cup C_{s}^{o}\left(A,\left(q_{s}^{L}\right)^{o}\right)\mid.
\]
 By Lemma 1 and the fact that $L$ satisfies either $\mid C_{s}^{o}\left(A,\left(q_{s}^{L}\right)^{o}\right)\cup A^{OBC}\mid=\left(q_{s}^{1}\right)^{o}+\left(q_{s}^{1}\right)^{OBC}$
or $\left(q_{s}^{L}\right)^{OP}=\left(q_{s}^{1}\right)^{o}+\left(q_{s}^{1}\right)^{OBC}$
in multi-run DA, we conclude 
\[
\mid A^{OBC}\cup C_{s}^{o}\left(A,\left(q_{s}^{n}\right)^{o}\right)\mid=\mid A^{OBC}\cup C_{s}^{o}\left(A,\left(q_{s}^{L}\right)^{o}\right)\mid.
\]
 Moreover, since $C_{s}^{o}$ is q-responsive, we obtain 
\[
A^{OBC}\cup C_{s}^{o}\left(A,\left(q_{s}^{n}\right)^{o}\right)=A^{OBC}\cup C_{s}^{o}\left(A,\left(q_{s}^{L}\right)^{o}\right),
\]
 which implies 
\[
A^{OBC}\cup C_{s}^{o}\left(A,\left(q_{s}^{L}\right)^{o}\right)\subseteq C_{s}^{BT}\left(A,q_{s}\right).
\]
 Moreover, by the construction of multi-run DA, we have 
\[
\begin{array}{c}
\mid A^{SC}\setminus C_{s}^{o}\left(A,\left(q_{s}^{L}\right)^{o}\right)\mid\leq\left(q_{s}^{1}\right)^{SC}\\
\mid A^{ST}\setminus C_{s}^{o}\left(A,\left(q_{s}^{L}\right)^{o}\right)\mid\leq\left(q_{s}^{1}\right)^{ST}
\end{array}
\]
because there are no de-reservations from categories SC and ST, and
hence, capacities of these categories remain unchanged in the course
of multi-run DA. Thus, we have 
\[
\begin{array}{c}
A^{SC}\setminus C_{s}^{o}\left(A,\left(q_{s}^{L}\right)^{o}\right)\subseteq C_{s}^{BT}\left(A,q_{s}\right)\\
A^{ST}\setminus C_{s}^{o}\left(A,\left(q_{s}^{L}\right)^{o}\right)\subseteq C_{s}^{BT}\left(A,q_{s}\right)
\end{array}
\]
 which completes our proof. Hence, the outcome of the multi-run DA
at preference profile $P$, i.e., the matching $v$, is individually
rational for every institution with respect to their backward transfers
choice rules. 

\paragraph{No Blocking. }

Toward a contradiction, suppose that $\left(i,s\right)$ is a blocking
pair. That is, $sP_{i}v(i)$ and $i\in C_{s}^{BT}(v(s)\cup\{i\})$.
Let $A_{s}^{SC}$, $A_{s}^{ST}$, and $A_{s}^{OBC}$ denote the set
of applicants in $v(s)$ that are members of SC, ST, and OBC, respectively.
There are three cases to consider. 
\begin{enumerate}
\item $t_{i}=OBC$. Since $sP_{i}v(i)$ and OBC applicants get weakly better
of in multi-run DA, we can conclude that individual $i$ applied to
$s$ and get rejected by $s$ in every iteration of the multi-run
DA. Since $i$ was never chosen, we know that the number of chosen
OBC members is at least as high as the initial capacity of the OBC
category in every iteration. Therefore, $\mid A_{s}^{OBC}\mid\geq q_{s}^{OBC}$.
Moreover, individual $i$ is not in top $q_{s}^{o}$ candidates in
the set $v(s)$ and every applicant in $A_{s}^{OBC}$ has higher merit
score than $i$. Thus, $i$ cannot be chosen from $v(s)\cup\{i\}$
in the backward transfers choice rule $C_{s}^{BT}$. This contradicts
with $(i,s)$ being a blocking pair. 
\item $t_{i}=SC$ or $t_{i}=ST$. Let us consider $t_{i}=SC$. Since $sP_{i}v(i)$
and SC applicants get weakly better of in multi-run DA, we can conclude
that individual $i$ applied to $s$ and get rejected by $s$ in every
step of the multi-run DA. Since $i$ is not chosen by $s$, all candidates
in $A_{s}^{GC}\cup A_{s}^{SC}$ have higher scores than $i$. Moreover,
there is no unfilled seat at reserved SC category. Since adding SC
candidates who have lower scores than candidates in $A_{s}^{GC}\cup A_{s}^{SC}$
to $v(s)$ cannot change the capacity vector of the final iteration
of $C_{s}^{BT}$, $i$ cannot be chosen from $v(s)\cup\{i\}$. This
is a contradiction. The case where $t_{i}=ST$ is proved similarly. 
\item $t_{i}=GC$. Since $sP_{i}v(i)$ and GC applicants get weakly better
of in multi-run DA, individual $i$ applied to $s$ and get rejected
by $s$ in every iteration of the multi-run DA. Since $i$ was never
chosen, we know that $i$ is not in top $\left(q_{s}^{N}\right)^{o}$,
i.e., number of open-category seats in the final iteration of the
multi-run DA. Therefore, $i$ cannot be chosen from $v(s)\cup\{i\}$.
This is a contradiction. Therefore, there is no blocking pair. 
\end{enumerate}
We have shown that the outcome of the multi-run DA at problem $P$,
i.e., matching $v$, is stable with respect to backward transfers
choice rules. Given that multi-run DA and DA-BT are different mechanisms,
individual-optimality of the DA-BT implies that it Pareto dominates
the multi-run DA. That is, the applicant-optimal stable (with respect
to backward transfers choice rules $C^{BT}$) matching $\mu$ Pareto
dominates matching $v$. This ends our proof. 

\paragraph{Proof of Theorem 4. }

We first show that the DA-BT mechanism satisfies all the axioms. Let
$\succ\equiv\left(\succ_{s}\right)_{s\in\mathcal{S}}$ be the profile
of institutional priorities and $C^{BT}=\left(C_{s}^{BT}\right)$
be the profile of institutions' BT choice rules. 

\paragraph{Individual rationality.}

In the course of the DA algorithm under BT choice rules, no individual
$i\in\mathcal{I}$ proposes to an unacceptable institution. Therefore,
the DA-BT mechanism is individually rational. 

\paragraph{Meritocracy.}

Given the profile of preferences and category memberships $\left(P,T\right)\in\mathcal{P}\times\mathcal{T}$
of individuals. Let $\eta$ be the outcome of the DA-BT mechanism
and $\mu$ be the matching induced by the assignment $\eta$. Consider
a pair $\left(i,s\right)\in\mathcal{I}\times\mathcal{S}$ such that
$sP_{i}\mu(i)$. Since individual $i$ is not matched with institution
$s$, $i$ must have applied to institution $s$ and get rejected
during the DA algorithm with respect to $\left(C_{s}^{BT}\right)_{s\in\mathcal{S}}$.
Let $n$ be the step at which individual $i$ gets rejected and $\mathcal{A}_{s}^{n}$
be the union of the set of individuals who were tentatively held by
institution $s$ at the end of Step $n-1$ and the set of new proposers
of $s$ in Step $n$. We have $i\in\mathcal{A}_{s}^{n}$ and $i\notin C_{s}^{BT}\left(\mathcal{A}_{s}^{n},q_{s}\right)$.
Let $N$ be the last iteration in the computation of $C_{s}^{BT}(\mathcal{A}_{s}^{n},q_{s}).$
By definition of $C_{s}^{BT}$, we have 
\[
C_{s}^{BT}(\mathcal{A}_{s}^{n},q_{s})=C_{s}^{IN}(\mathcal{A}_{s}^{n},q_{s}^{N}).
\]

Since $i\notin C_{s}^{IN}\left(\mathcal{A}_{s}^{n},q_{s}^{N}\right)$,
for all $\left(j,o\right)\in\eta(s)$ we have $j\succ_{s}i$. If $t(i)\in\mathcal{R}$,
then for all $\left(j,t(i)\right)\in\eta(s)$ , we have $j\succ_{s}i$,
by the definition of $C_{s}^{IN}$. 

\paragraph{Non-wastefulness. }

In the DA algorithm with respect to $\left(C_{s}^{BT}\right)_{s\in\mathcal{S}}$,
any applicant $i$ is rejected from open-category at institution $s$
if $q_{s}^{o}+q_{s}^{OBC}$ positions are exhausted in the last iteration
of $C_{s}^{BT}$ in the step of the DA where $i$ is rejected. Similarly,
an applicant $i$ with $t\left(i\right)=r$ is rejected from reserve
category $r$ at institution $s$ if (1) she is rejected from open-category
in the last iteration of $C_{s}^{BT}$ which requires $q_{s}^{o}+q_{s}^{OBC}$
positions to be exhausted, and (2) $q_{s}^{t(i)}$ category $r$ positions
are exhausted in the last iteration of $C_{s}^{BT}$, at the step
of the DA where $i$ is rejected. Hence, the DA-BT mechanism is non-wasteful. 

\paragraph{Open-first. }

Given the profile of preferences and category memberships $\left(P,T\right)\in\mathcal{P}\times\mathcal{T}$
of individuals, let $n$ be the \emph{last }step of the DA with respect
to $\left(C_{s}^{BT}\right)_{s\in\mathcal{S}}$. Let $\mathcal{A}_{s}^{n}$
be the union of the set of individuals who were tentatively held by
institution $s$ at the end of Step $n-1$ and the set of new proposers
of $s$ in Step $n$. By definition of BT choice rule, $C_{s}^{BT}\left(\mathcal{A}_{s}^{n},q_{s}\right)=C_{s}^{IN}\left(\mathcal{A}_{s}^{n},q_{s}^{N_{s}}\right)=\mu_{s}$,
where $\mu_{s}$ is the set of individuals who are matched to institution
$s$, $N_{s}$ is the last iteration of $C_{s}^{BT}$ and $q_{s}^{N_{s}}$
is the distribution of positions over categories. Note that if $\mid\mathcal{A}_{s}^{n}\mid\leq q_{s}^{o}+q_{s}^{OBC}$
then all individuals in $\mathcal{A}_{s}^{n}$ is assigned either
open-category or OBC positions since all unfilled OBC positions are
reverted to open-category by the last iteration $N_{s}$. When $\mid\mathcal{A}_{s}^{n}\mid>q_{s}^{o}+q_{s}^{OBC}$,
by definition of $C_{s}^{BT}$, the number of individuals who are
assigned to either open-category or OBC positions is $q_{s}^{o}+q_{s}^{OBC}$.
Moreover, by definition of $C_{s}^{BT}$, each individual who receive
an open-category position has higher priority than any individual
who is assigned to a reserve category position. 

Let $\eta\left(s\right)$ denote the assignment of institution $s$.
For each individual $i\in C_{s}^{o}\left(\mathcal{A}_{s}^{n},(q_{s}^{N_{s}})^{o}\right)$,
$\left(i,o\right)\in\eta\left(s\right)$. Let $A^{'}=\mathcal{A}_{s}^{n}\setminus C_{s}^{o}\left(\mathcal{A}_{s}^{n},(q_{s}^{N_{s}})^{o}\right)$.
Then, for each $r\in\mathcal{R}$, $C_{s}^{r}\left(A^{'},\left(q_{s}^{N_{s}}\right)^{r}\right)$
denotes the set of applicants who are matched to category $r$ positions
in institution $s$. Take any $i$ with $\left(i,o\right)\in\eta\left(s\right)$
and $t_{i}=r$. Consider the assignment $\eta^{'}\left(s\right)=\eta\left(s\right)\setminus\left\{ \left(i,o\right)\right\} \cup\left\{ \left(i,r\right)\right\} $.
If $i$ is not the lowest-ranked individual among $C_{s}^{o}\left(\mathcal{A}_{s}^{n},(q_{s}^{N_{s}})^{o}\right)$,
then removing $\left(i,o\right)$ and adding $\left(i,r\right)$ violates
that each applicant in open-category position has higher priority
than any individual who is assigned to a reserve category position.
Now suppose $i$ is the lowest-ranked applicant in $C_{s}^{o}\left(\mathcal{A}_{s}^{n},(q_{s}^{N_{s}})^{o}\right)$.
Notice that $i$ cannot be a member of OBC because otherwise that
last position would not be reverted from OBC to open-category. Hence,
$i$ must be either $SC$ or $ST$ member. In both cases, removing
$\left(i,o\right)$ and adding $\left(i,r\right)$ violates that the
number of individuals who are assigned to either open-category or
OBC positions is $q_{s}^{o}+q_{s}^{OBC}$. This ends our proof. 

\paragraph{Incentive-compatibility.}

We have shown incentive-compatibility in the proof of Theorem 3. 

We show uniqueness with two lemmata. 
\begin{lem}
Let $\varphi$ be a mechanism that is individually rational, non-wasteful,
and satisfies meritocracy and open-first. Let $\left(P,T\right)\in\mathcal{P}\times\mathcal{T}$,
$\varphi\left(P,T\right)=\eta$, and $\mu=\mu\left(\eta\right)$.
Then, $\mu$ is \emph{stable} with respect to the profile of BT choice
rules $\mathbf{C}^{BT}=\left(C_{s}^{BT}\right)_{s\in\mathcal{S}}$,
i.e., 

(1) for every individual $i\in\mathcal{I}$, $\mu_{i}\left(\eta\right)R\emptyset$, 

(2) for every institution $s\in\mathcal{S}$, $C_{s}^{BT}(\mu_{s},q_{s})=\mu_{s}$,
and 

(3) there is no $(i,s)$ such that $sP_{i}\mu_{i}$ and $i\in C_{s}^{BT}(\mu_{s}\cup\{i\},q_{s})$.
\end{lem}

\paragraph{Proof of (1). }

By individual rationality of $\varphi$ the assignment $\eta=\varphi\left(P,T\right)$
is individually rational. Hence, condition (1) is trivially satisfied. 

\paragraph{Proof of (2). }

Consider any $s\in\mathcal{S}$. We need to show that $C_{s}^{BT}(\mu_{s},q_{s})=\mu_{s}$.
Towards a contradiction, suppose that $C_{s}^{BT}(\mu_{s},q_{s})\neq\mu_{s}$.
Then, there exists an individual $i$ with $\left(i,c\right)\in\eta(s)$
for some $c\in\mathcal{C}$ such that $i\notin C_{s}^{BT}\left(\mu_{s},q_{s}\right)$.
That means 
\begin{itemize}
\item $\left(q_{s}^{o}+q_{s}^{OBC}\right)$ positions are allocated either
as open-category or OBC category positions, by definition of $C_{s}^{BT}$,
and
\item individual $i$ has a lower ranking than every individual who receives
an $t_{i}$ category positions (if $t_{i}\in\mathcal{R}$) and open-category
position (either from the initially set $q_{s}^{o}$ open-category
positions or reverted unfilled OBC positions).
\end{itemize}
Since $\eta$ satisfies open-first and $\left(q_{s}^{o}+q_{s}^{OBC}\right)$
positions are allocated either as open-category or OBC category positions
under $C_{s}^{BT}$, we have 
\[
\mid\left\{ j\mid\left(j,o\right)\in\eta(s)\;or\;(j,OBC)\in\eta(s)\right\} \mid=q_{s}^{o}+q_{s}^{OBC}.
\]

\subparagraph{Case 1. }

Suppose $t_{i}=OBC$. $i\notin C_{s}^{BT}\left(\mu_{s},q_{s}\right)$
implies that all $q_{s}^{OBC}$ positions are taken by higher priority
OBC individuals. That means there is no de-reservation and \emph{at
least} $q_{s}^{OBC}$ individuals---who are members of OBC---are
selected in open-category and OBC category. For all $j\in C_{s}^{BT}\left(\mu_{s},q_{s}\right)$
such that $t_{j}=OBC$, we have $\left(j,o\right)\in\eta(s)$ or $\left(j,OBC\right)\in\eta\left(s\right)$
since $\eta$ satisfies open-first. Let $x=\mid\left\{ j:t_{j}=OBC,\;and\;j\in C_{s}^{BT}\left(\mu_{s},q_{s}\right)\right\} \mid$
be the number of OBC members chosen under $C_{s}^{BT}\left(\mu_{s},q_{s}\right)$.
Open-first implies that, under assignment $\eta$, top $\left(q_{s}^{o}+q_{s}^{OBC}-x\right)$
individuals with respect to $\succ_{s}$ must be assigned an open-category
position. Similarly, under $C_{s}^{BT},$ top $\left(q_{s}^{o}+q_{s}^{OBC}-x\right)$
individuals with respect to $\succ_{s}$ are selected from open-category
in the last step of $C_{s}^{BT}$. Note that each individual who are
selected from open-category under $C_{s}^{BT}$ are assigned an open-category
position under $\eta$. Moreover, each individual who are selected
from OBC category under $C_{s}^{BT}$ are assigned either open-category
or OBC position under $\eta$ since $\eta$ satisfies open-first.
Hence, $i$ cannot be assigned under $\eta$. Otherwise, it violates
open-first. Therefore, no member of OBC can be assigned under $\eta$
and not selected under $C_{s}^{BT}$. This is a contradiction. 

\subparagraph{Case 2. }

Suppose $t_{i}=GC$. Each individual who receives an OBC position
under $\eta$ is selected under $C_{s}^{BT}$ and $\eta$ satisfies
open-first imply that each individual in top $\left(q_{s}^{o}+q_{s}^{OBC}-x\right)$
individuals with respect to $\succ_{s}$ are assigned open-category
positions under $\eta$. By definition of $C_{s}^{BT}$, $\left(q_{s}^{o}+q_{s}^{OBC}-x\right)$
positions are allocated as open-category positions. Therefore, each
individual who receives either open-category or OBC positions must
be selected under $C_{s}^{BT}$. This is a contradiction. 

\subparagraph{Case 3. }

Suppose $t_{i}=SC$. Since $i\notin C_{s}^{BT}\left(\mu_{s},q_{s}\right)$
we obtain $i$ has a lower priority than each individual who are selected
from open-category in the last step of $C_{s}^{BT}$ and has a lower
priority than every individual who are selected in SC category under
$C_{s}^{BT}$. As we stated in Case 2, each individual who receives
either open-category or OBC positions must be selected under $C_{s}^{BT}$.
This implies that $\left(i,o\right)\notin\eta\left(s\right)$. The
only other possibility is that $\left(i,SC\right)\in\eta\left(s\right)$.
This contradicts with the feasibility of $\eta$. Similarly, $t_{i}=ST$
cannot be the case. A similar proof applies for this case. 

\paragraph{Proof of (3).}

Consider a pair $\left(i,s\right)\in\mathcal{I}\times\mathcal{S}$
such that $sP_{i}\mu_{i}$. Consider the case where $t(i)\in\left\{ SC,ST\right\} $.
Without loss of generality, let $t(i)=SC$. Let $\eta=\varphi(P,T)$
and $\mu=\mu\left(\eta\right)$. By non-wastefulness of $\eta$, we
have $\mid\left\{ j\mid\left(j,SC\right)\in\eta(s)\right\} \mid=q_{s}^{SC}$.
Also, by meritocracy of $\eta$, for all $j$ such that $\left(j,SC\right)\in\eta(s)$,
we must have $j\succ_{s}i$. Therefore, individual $i$ cannot be
chosen from the category SC under $C_{s}^{BT}$. Moreover, since $\eta$
satisfies open-first, we obtain 
\[
j^{'}\succ_{s}j\succ_{s}i
\]
 for all $j$ and $j^{'}$ such that $\left(j^{'},o\right)\in\eta(s)$
and $\left(j,SC\right)\in\eta(s)$. Hence, $i\notin C_{s}^{BT}\left(\mu_{s}\cup\{i\},q_{s}\right)$.
Note that adding $i$ to the set $\mu_{s}$ does not change the number
of OBC positions that are reverted to open-category. 

Suppose $t(i)=OBC.$ By non-wastefulness of $\eta$ we have 
\[
\mid\left\{ j\mid\left(j,OBC\right)\in\eta(s)\right\} \mid=q_{s}^{OBC}.
\]
There is no vacant OBC position in institution $s$ under $\eta$.
Hence, no position is reverted from OBC to open-category. By meritocracy
of $\eta$, for all $j$ such that $\left(j,OBC\right)\in\eta(s)$,
we have $j\succ_{s}i$. $\eta$ satisfying open-first implies 
\[
j^{'}\succ_{s}j\succ_{s}i
\]
 for all $j$ and $j^{'}$ such that $\left(j^{'},o\right)\in\eta(s)$
and $\left(j,OBC\right)\in\eta(s)$. This means $i\notin C_{s}^{BT}\left(\mu_{s}\cup\{i\},q_{s}\right)$
since all individuals who receive either open-category or OBC positions
have higher priorities than $i$ at $s$.

Finally, consider $t(i)=GC$. In this case, non-wastefulness of $\eta$
implies 
\[
\mid\left\{ j\mid\left(j,o\right)\in\eta(s)\;or\;(j,OBC)\in\eta(s)\right\} \mid=q_{s}^{o}+q_{s}^{OBC}.
\]

Meritocracy of $\eta$ implies that $j\succ_{s}i$ for all $j$ such
that $\left(j,o\right)\in\eta(s)$. Adding $i$ to $\mu_{s}$ does
not change the number of OBC positions to be reverted to open-category.
We conclude that $i\notin C_{s}^{BT}\left(\mu_{s}\cup\{i\},q_{s}\right)$.
Thus, independent of individual $i$'s category membership, the pair
$\left(i,s\right)$ cannot be a blocking pair, which ends our proof. 
\begin{lem}
The DA-BT mechanism is the only mechanism that is stable and strategy-proof. 
\end{lem}

\paragraph{Proof.}

Incentive-compatibility of $\varphi$ implies that $\varphi$ is strategy-proof.
We show in Lemma 2 that $\varphi$ is a stable mechanism where stability
is defined with respect to $C=\left(C_{s}^{BT}\right)_{s\in\mathcal{S}}$. 

We have already shown that $C_{s}^{BT}$ is substitutable and size
monotonic. Therefore, by Hatfield, Kominers, and Westkamp (2021),
$C_{s}^{BT}$ satisfies observable substitutability, observable size
monotonicity, and non-manipulability via contractual terms conditions.
By Theorem 4 of Hatfield, Kominers, and Westkamp (2021), DA-BT is
the unique stable and strategy-proof mechanism. Therefore, $\varphi=DA-BT$.
Note that incentive-compatibility also requires that given a preference
profile of individuals no reserve category member can get better of
by not revealing their reserve category membership. We show in the
proof of Theorem 3 that DA-BT cannot be manipulable via not revealing
reserve category memberships. 
\end{document}